\documentclass[12pt]{article}

\usepackage{latexsym}
\usepackage{amssymb,amsfonts,amsmath}
\usepackage{graphicx} 
\usepackage{indentfirst}
\usepackage{bbm}
\usepackage{amssymb}
\usepackage{verbatim}
\usepackage{amsmath, amsthm,amssymb}
\usepackage{mathrsfs}
\usepackage{hyperref}
\usepackage{amsfonts}
\usepackage{dsfont}
\usepackage{cite}
\usepackage{xcolor}

\parskip=\medskipamount

\newcommand {\cC}{{\cal C}}
\newcommand {\cD}{{\cal D}}

\newcommand {\cM}{{\cal M}}
\newcommand {\cN}{{\cal N}}
\newcommand {\cO}{{\cal O}}

\newcommand {\cQ}{{\cal Q}}

\newcommand {\cS}{{\cal S}}

\newcommand {\cW}{{\cal W}}

\def\a{\alpha}

\def\b{\beta}

\def\d{\delta}
\def\e{\epsilon}

\def\g{\gamma}

\def\l{\lambda}

\def\q{\theta}
\def\r{\rho}
\def\s{\sigma}

\def\x{\xi}
\def\z{\zeta}

\def\F{\Phi}
\def\J{\Psi}
\def\L{\Lambda}
\def\O{\Omega}

\def\S{\Sigma}
\def\U{\Upsilon}

\def\rd{{\rm d}}
\def\ri{{\rm i}}
\def\re{{\rm e}}

\newcommand{\ve}{\varepsilon}                            

\newcommand{\pa}{\partial}                           
\newcommand{\hf}{\frac12}

\newcommand{\vf}{\varphi}

\newcommand{\be}{\begin{equation}}
\newcommand{\ee}{\end{equation}}
\newcommand{\bea}{\begin{eqnarray}}
\newcommand{\eea}{\end{eqnarray}}
\newcommand{\non}{\nonumber}
\newcommand{\1}{{\underline{1}}}

\newcommand{\bm}[1]{\mbox{\boldmath$#1$}}

\def\double #1{#1{\hbox{\kern-2pt $#1$}}}

\newif\ifdtup

\newcommand{\bsubeq}{\begin{subequations}}
\newcommand{\esubeq}{\end{subequations}}

\numberwithin{equation}{section}

\newcommand{\sSp}{\mathsf{Sp}}
\newcommand{\sSU}{\mathsf{SU}}
\newcommand{\sSL}{\mathsf{SL}}
\newcommand{\sGL}{\mathsf{GL}}
\newcommand{\sSO}{\mathsf{SO}}

\newcommand{\sOSp}{\mathsf{OSp}}

\newcommand{\mc}{\mathcal}
\newcommand{\mf}{\mathfrak}

\newcommand{\mb}{\mathbb}

\begin{document}

\begin{titlepage}
\begin{flushright}
March, 2021 \\
\end{flushright}
\vspace{5mm}

\begin{center}
{\Large \bf 

Higher-spin Cotton tensors and massive gauge-invariant actions in AdS$_3$ 
}
\end{center}

\begin{center}

{\bf Sergei M. Kuzenko and Michael Ponds} \\
\vspace{5mm}

\footnotesize{
{\it Department of Physics M013, The University of Western Australia\\
35 Stirling Highway, Perth W.A. 6009, Australia}}  
~\\
\vspace{2mm}
~\\
Email: \texttt{ 
sergei.kuzenko@uwa.edu.au, michael.ponds@research.uwa.edu.au}\\
\vspace{2mm}

\end{center}

\begin{abstract}
\baselineskip=14pt
In a conformally flat three-dimensional spacetime, 
the linearised higher-spin Cotton tensor $\mathfrak{C}_{\a(n)}(h)$  is the unique 
conserved conformal current which is 
a gauge-invariant descendant of the conformal gauge prepotential $h_{\a(n)}$.
The explicit form of $\mathfrak{C}_{\a(n)}(h)$ is well known in Minkowski space. Here we solve the problem of extending the Minkowskian result to the case of anti-de Sitter (AdS) space and derive a closed-form expression for 
$\mathfrak{C}_{\a(n)}(h)$ in terms of the AdS Lorentz covariant derivatives.
It is shown that every conformal higher-spin action 
$S_{\text{CS}}^{(n)}[h]\propto \int\text{d}^3x\, e \, h^{\a(n)}\mf{C}_{\a(n)}(h) $ factorises into a product of $(n-1)$ first-order operators that are associated with the spin-$n/2$ partially massless  AdS values. 
Our findings greatly facilitate the on-shell analysis of massive higher-spin gauge-invariant actions in AdS$_3$. 
The main results are extended to the case of $\cN=1$ AdS supersymmetry. In particular, we derive simple expressions for the higher-spin super-Cotton tensors in AdS$_3$.  
\end{abstract}
\vspace{5mm}

\vfill

\vfill
\end{titlepage}

\newpage
\renewcommand{\thefootnote}{\arabic{footnote}}
\setcounter{footnote}{0}

\tableofcontents{}
\vspace{1cm}
\bigskip\hrule

\allowdisplaybreaks

\section{Introduction}

The conformal geometry of spacetime in four and higher dimensions is controlled by
the Weyl tensor. The necessary and sufficient condition for spacetime to be conformally flat is that the Weyl tensor is equal to zero, see e.g.  \cite{Eisen}.
If conformal gravity is realised as the gauge theory of the conformal group \cite{KTvN}, 
then the algebra of conformally covariant derivatives is determined by the Weyl tensor and 
its covariant derivatives \cite{ButterN=1,BKNT-M1,BKNT}. The conformal connection becomes flat if the Weyl tensor vanishes.

In three dimensions, the Weyl tensor is identically zero and all information 
about conformal geometry is encoded in the Cotton tensor, $C_{abc}=- C_{bac}$. 
Spacetime is conformally flat if and only if the Cotton tensor vanishes \cite{Eisen}
(see \cite{BKNT-M1} for a modern proof).  The commutator of two conformally covariant derivatives $\nabla_{a}$ involves only the Cotton tensor \cite{BKNT-M1} (see also \cite{KP19})
\begin{subequations}\label{1.1}
\bea
\big[\nabla_a,\nabla_b\big] =\frac{1}{2}C_{abc}K^c~, \qquad C_{abc}=-\ve_{abd}C^d{}_{c}~,
\label{1.1a}
\eea
where $K^c$ is the special conformal generator.
The algebraic structure of the Cotton tensor is described by the relations
\bea
 C_{ab}=C_{ba}~,\qquad C^b{}_{b}=0~.
\eea
 The Cotton tensor is a primary field, $K^c C_{ab}=0$, 
  of dimension $3$ and obeys the Bianchi identity
\bea
\nabla^b C_{ba} =0~.
\eea
\end{subequations}
As follows from \eqref{1.1a},  
the conformal connection becomes flat if the Cotton tensor vanishes.

Since the pioneering work by Fradkin and Tseytlin \cite{FT}, there has been much interest in conformal higher-spin  (CHS) theories in diverse dimensions, see
\cite{FL-3D,PopeTownsend,FL,FL-4D,Tseytlin,Segal,Marnelius,Metsaev10,Vasiliev2009,BJM2}
for an incomplete list of works published within a quarter-century after \cite{FT}.
This interest has truly exploded in the last decade (see, e.g., \cite{BG,Metsaev:2014iwa,NTCHS,Beccaria:2014jxa,Beccaria:2016syk,GrigorievT,KMT,BeccariaT,Bonezzi,Manvelyan2,Adamo:2018srx,Grigoriev:2019xmp,Grigoriev:2020lzu} and references therein), and  some comments 
 on the literature will be given below. In the case of three dimensions, higher-spin generalisations of the Cotton tensor are of central importance and lie at the heart of many recent studies \cite{BHT,BKRTY,Nilsson1, Nilsson2,HHL, LN, KO, K16, KT, BBB,KP18,HLLMP,BHHK}.  
 Linearised higher-spin Cotton tensors in AdS$_3$ will be the main object of interest in this work.

In Minkowski space, linearised higher-spin extensions
of the Cotton tensor were constructed in the bosonic
\cite{PopeTownsend} and fermionic \cite{K16}  cases.
An alternative derivation of the higher-spin Cotton tensors
was given in \cite{HHL,HLLMP}, where the latter played an integral role in establishing a conformal geometry of higher-spin gauge fields. Supersymmetric extensions of the higher-spin Cotton tensors were derived in 
\cite{K16,KT} for $\cN=1$, in \cite{KO} for $\mc{N}=2$ and in \cite{BHHK} for  $\cN>2$ Poincar\'e supersymmetry. 
It was argued in \cite{KP18,KP19} that 
higher-spin Cotton tensors exist on any conformally flat background, although explicit expressions for them in terms of Lorentz covariant derivatives and conformal prepotentials are difficult to derive if the spacetime curvature is non-vanishing. It is pertinent here  to elaborate on these points in some more detail.

Given a conformally flat spacetime $\cM^3$, its geometry may be described in terms of 
torsion-free Lorentz covariant derivatives 
\begin{align}
\mc{D}_{a}
=e_a+\omega_a
=e_a{}^{m}\pa_m+\frac{1}{2}\omega_{a}{}^{bc}M_{bc}
~, \qquad \big[ \cD_a , \cD_b \big] = \hf R_{ab}{}^{cd} M_{cd}~, \label{VecDer}
\end{align}
where $e_{a}{}^{m}$ is the inverse vielbein, $\omega_{a}{}^{bc}$  the Lorentz connection, 
and $M_{ab} =-M_{ba}$  the Lorentz generators. Since $\cM^3$ is conformally flat, its Cotton tensor $C_{abc}$ is zero, 
\bea
C_{abc} := \cD_a P_{bc} -\cD_b P_{ac} =0~, \qquad 
P_{ab} = R_{ab} -\frac 14 \eta_{ab} R~,
\label{Cotton33}
\eea
where $P_{ab}$ denotes the Schouten tensor.  
Let $\z = \z^m \pa_m = \z^a e_a$ 
be a conformal Killing vector field on $\cM^3$, 
\bea
 \cD^a \z^b + \cD^b \z^a = 2 \eta^{ab} \s[\z] ~, \qquad \s[\z] = \frac{1}{3} \cD_b \z^b  ~.
\label{CCVF1.3}
 \eea
 The conformal Killing vector fields 
span  the conformal algebra of $\cM^3$, which is isomorphic to $\mathfrak{so}(3,2)$. A primary tensor field $\vf$ (with suppressed indices) possesses the following conformal transformation law
\bea
-\d_\z \vf = \Big(\z^b \cD_b +\hf K^{bc}[\z] M_{bc} + d_\vf \s[\z]\Big) \vf ~, \qquad 
K^{bc} [\z] &=& \hf \big( \cD^b \z^c - \cD^c \z^b \big)~, \label{ConfT}
\eea
where $d_\vf$ is the dimension of $\vf$. 

Let us fix a positive integer $n \geq 2$.
In the two-component spinor notation, the conformal  gauge field $h_{\a(n)} :=h_{\a_1 \dots \a_n} = h_{(\a_1 \dots \a_n)}$ is a real primary field of dimension 
\bea
d_{h_{\a(n)}} = 2-\frac{n}{2}  \label{1.7}
\eea
with the gauge transformation law\footnote{Here, and often below, we  make use of the compact notational convention introduced by Vasiliev 
\cite{Vasiliev87}: 
$V_{ \a(n)} U_{\a(m) }= V_{ (\a_1 \dots \a_n} U_{\a_{n+1} \dots \a_{n+m})}$.
}
\begin{align}
\delta_{\x}h_{\a(n)}=\mc{D}_{\a(2)}\x_{\a(n-2)}~.\label{GT}
\end{align}
The Cotton tensor $\mf{C}_{\a(n)}(h)$ associated with  $h_{\a(n)}$ 
is defined to satisfy the following three conditions: 
(i) it is a primary descendent of $h_{\a(n)}$; (ii) it is transverse (or covariantly conserved),
\begin{align}
0=\mc{D}^{\b(2)}\mf{C}_{\b(2)\a(n-2)}(h) \label{cotT}~;
\end{align}
(iii) it is invariant under the gauge transformations \eqref{GT},
\begin{align}
0=\mf{C}_{\a(n)}(\delta_{\x }h)~. \label{cotG}
\end{align}
The conditions (i) and (ii)    
fix the dimension of $\mf{C}_{\a(n)}(h)$
to be
\bea
d_{\mf{C}_{\a(n)}} = 1 +\frac{n}{2} ~.
\label{1.9}
\eea
The Cotton tensor $\mf{C}_{\a(n)}(h)$ proves to exist, and is unique
modulo an overall normalisation. The reader can ask the natural question: What goes wrong if the background manifold is not conformally flat?
The answer is that the properties \eqref{cotT} and \eqref{cotG} break down for $n>2$, 
\bea
C_{abc}\neq 0 \quad \implies \quad 
\mc{D}^{\b(2)}\mf{C}_{\b(2)\a(n-2)}(h) = \cO(C) ~, \qquad 
\mf{C}_{\a(n)}(\delta_{\x }h) =\cO(C)~.
\eea
We refer the reader to \cite{KP18, KP19} for the technical details. 

In Minkowski space, the covariant derivative can be chosen to coincide with a partial one, $\cD_a = \d_a{}^m \pa_m = \pa_a$. Then one can derive the following closed-form expression for the higher-spin Cotton tensor\footnote{Relation \eqref{Mcot} was obtained 
in \cite{K16} via the component reduction of the  higher-spin $\cN=1$ super-Cotton tensor, eq. \eqref{5.17}, which is remarkably simple. This is an example of the power of supersymmetry.}
 \cite{K16} 
\begin{align}
{\mathfrak C}_{\a(n)}(h)&:=\frac{1}{2^{n-1}} \sum\limits_{j=0}^{  \lceil {n/2} \rceil -1 }\binom{n}{2j+1}\Box^{j}\big(\pa_{\a}{}^{\b}\big)^{n-2j-1}h_{\a(2j+1)\b(n-2j-1) }
\label{Mcot}~.
\end{align}
In the bosonic case, $n = 2s$, where $s\geq 1$ is an integer, the above expression is equivalent to those given in \cite{PopeTownsend,HHL}. 
It should be pointed out that the conformal spin-3 case, $n=6$, 
was studied for the first time in \cite{DD}.
 The spin-3/2 case, $n=3$,  
was considered in \cite{ABdeRST}. 
The field strength $\mf{C}_{\a(3)}(h)$ is the linearised version of the
Cottino vector-spinor \cite{DK,GPS}.

The extension of \eqref{Mcot} to a conformally flat spacetime $\cM^3$ is obtained \cite{KP19} by replacing $\pa_a \to \nabla_a$. Since the commutator \eqref{1.1a} vanishes in every conformally flat spacetime, all the properties of ${\mathfrak C}_{\a(n)}(h)$ follow through.
However, in addition to the Lorentz generators, $\nabla_a$ involves the special conformal generators $K_b$  (in a special gauge where the dilatation connection is gauged away).
In order to switch from the description in terms of $\nabla_a$ to that of the standard Lorentz covariant derivative $\cD_a$, one has to go through the process of degauging 
\cite{BKNT-M1,KP19}. This includes expanding $\nabla_{a}$ using the relation
\bea
\nabla_a=\mathcal{D}_a+\frac 12 P_{a}{}^{b}K_b
\eea 
and then evaluating the action of all special conformal generators.
The description of ${\mathfrak C}_{\a(n)}(h)$ in terms of $\nabla_a$ suffices to describe 
the conformal higher-spin action \eqref{CSA}. However, in order to formulate non-conformal massive actions such as \eqref{HSNTMG}, one has to express 
${\mathfrak C}_{\a(n)}(h)$ in terms of the Lorentz covariant derivatives $\cD_a$.

It turns out that, even in the case of a maximally symmetric spacetime, the implementation of the degauging procedure for the higher-spin Cotton tensor is extremely non-trivial. For example, in AdS$_3$ the expressions for $\mf{C}_{\a(n)}(h)$ with $3\leq n \leq 6$ are given by \cite{KP18}
\begin{subequations}\label{LowCot}
\begin{align}
\mathfrak{C}_{\a(3)}(h)=&\frac{1}{2^2}\bigg(3\big(\cD_{\a}{}^{\b}\big)^2
{h}_{\a\b(2)}+\big(\cQ -9\mc{S}^2\big){h}_{\a(3)}\bigg)~,\\
\mathfrak{C}_{\a(4)}(h)=&\frac{1}{2^3}\bigg(4\big(\cD_{\a}{}^{\b}\big)^3h_{\a\b(3)}
+4( \cQ  - 20 \cS^2) \cD_{\a}{}^{\b}{h}_{\a(3)\b}
\bigg)~,\\
\mathfrak{C}_{\a(5)}(h)=&\frac{1}{2^4}\bigg(5\big(\cD_{\a}{}^{\b}\big)^4{h}_{\a\b(4)}
+10\big(\cQ-33\mc{S}^2\big)\big(\cD_{\a}{}^{\b}\big)^2{h}_{\a(3)\b(2)}
\notag\\
&\phantom{extra}
+\big(\cQ-57\mc{S}^2\big)\big(\mc{Q}-25\mc{S}^2\big) {h}_{\a(5)}\bigg)~,\\
\mathfrak{C}_{\a(6)}(h)=&\frac{1}{2^5}\bigg(6\big(\cD_{\a}{}^{\b}\big)^5{h}_{\a\b(5)}
+20\big(\cQ-48\mc{S}^2\big)\big(\cD_{\a}{}^{\b}\big)^3{h}_{\a(3)\b(3)}\notag\\
&\phantom{extra}+\big(6\cQ^2-704\mathcal{S}^2+18432\mathcal{S}^4\cQ\big)\cD_{\a}{}^{\b}h_{\a(5)\b}\bigg)~,
\end{align}
\end{subequations}
where $\cQ$ is the second quadratic Casimir of the AdS group $\sSO(2,2)$,
given by eq. \eqref{Q}. No simple systematic method was offered in \cite{KP19} to degauge $\mf{C}_{\a(n)}(h)$ for generic $n$. 
In this paper we solve the problem of deriving a closed-form expression for 
$\mathfrak{C}_{\a(n)}(h)$ in terms of the AdS Lorentz covariant derivatives. An analogous result is also derived in the case of $\mc{N}=1$ AdS$_3$ supersymmetry. 

This paper is organised as follows. In section \ref{section 2} we begin by reviewing 
the structure of on-shell partially massless and massive fields in AdS$_3$. The Cotton tensors $\mf{C}_{\a(n)}(h)$, for arbitrary $n\geq 2$, are then constructed and their factorisation into second (and first) order operators is discussed. In section \ref{section 3} we use the results of section \ref{section 2} to analyse the on-shell dynamics of (new) topologically massive higher-spin gauge models in AdS$_3$. Section \ref{section 4} is devoted to $\mc{N}=1$ supersymmetric extensions of the aforementioned results.     
Concluding remarks are given in section \ref{section 5}. The main body is accompanied by three technical appendices. Appendix \ref{appendix 0}  lists our two-component spinor conventions. Appendix \ref{appendix A} describes the generating function formalism, which is the framework used to derive many of the results in this work. 
In appendix \ref{AppendixC} we discuss some specific features of the higher-spin Cotton tensors in conformally flat backgrounds.


\section{Higher-spin Cotton tensors  and on-shell fields in AdS$_3$}\label{section 2}

In this section we present a closed-form expression for the higher-spin Cotton tensor in AdS$_3$ and give several applications of this result. Throughout this work, we use the two-component spinor formalism almost exclusively. The relevant conventions adopted in this paper are summarised in appendix \ref{appendix 0}.

\subsection{AdS$_3$ geometry} 

 The covariant derivatives of AdS$_3$ satisfy the commutation relations
\begin{align}
\big[\mc{D}_a,\mc{D}_b\big]=-4\mc{S}^2M_{ab}\qquad\Longleftrightarrow \qquad
\big[\mc{D}_{\a\b},\mc{D}_{\g\d}\big]=4\mc{S}^2\big(\ve_{\g(\a}M_{\b)\d}+\ve_{\d(\a}M_{\b)\g}\big) ~.
\label{alg2.1}
\end{align}
Here the parameter $\mc{S}$ is related to the AdS radius $\ell$ and
the scalar curvature $R$ via $\ell^{-1} = 2 \cS$ and  
$R=-24\mc{S}^2$, respectively. The Lorentz generators 
with vector ($M_{ab} =-M_{ba}$) and spinor  ($M_{\a\b}=M_{\b\a}$) indices 
are defined to act on a vector $V_c$ and a spinor $\Psi_{\g}$ as in 
\eqref{generators}.\footnote{We will also make use of the Lorentz generator with a vector index, $M_a$. The three types of generators $M_a$, $M_{ab}$ and $M_{\a\b}$  are related to each other as follows: 
$-K^aM_a=
\hf K^{ab}M_{ab}=\hf K^{\a\b}M_{\a\b}$. These relations follow the general rule \eqref{A.7}.}
We parametrise the curvature in terms of $\cS$ in order for the notation to be consistent with that used in $\cN=1$ supergravity \eqref{N=1alg} and AdS superspace
\eqref{algN1}.

Isometries of AdS$_3$ are generated by those solutions of the conformal Killing equation \eqref{CCVF1.3}, which obey the additional restriction $\s[\z]=0$. 
They are called the Killing vector fields on AdS$_3$. Given a  tensor field $\vf$ on AdS$_3$, its AdS transformation law is obtained from \eqref{ConfT} by setting $\s[\z]=0$.  

There are two quadratic Casimir operators of the AdS$_3$ isometry algebra 
$\mf{so}(2,2)\cong \mf{sl}(2,{\mathbb R}) \oplus \mf{sl}(2,{\mathbb R})$, and they can be chosen as follows  (see, e.g., \cite{BPSS}):
\begin{subequations}
\bea
\mc{F}&:=&\mc{D}^{\a\b}M_{\a\b}~,\qquad ~~~~~~~~~~\phantom{..}\big[\mc{F}, \mc{D}_{\a(2)}\big]=0~, \label{F}
\\
\mc{Q}&:=&\Box - 2\mc{S}^2M^{\a\b}M_{\a\b}~,\qquad \big[\mc{Q},\mc{D}_{\a(2)}\big]=0~, \label{Q}
\eea
\end{subequations}
where $\Box= \mc{D}^a\mc{D}_{a}=-\frac{1}{2}\mc{D}^{\a\b}\mc{D}_{\a\b}$. These Casimir operators are independent of one another, 
which may be seen by the action of $\mc{F}^2$ on an  
unconstrained 
completely 
symmetric rank-$n$ spinor field $h_{\a(n)}=h_{(\a_1\dots\a_n)}$:
\begin{align}
\mc{F}^2h_{\a(n)}=n^2\big[\mc{Q}-(n-2)(n+2)\mc{S}^2\big]h_{\a(n)}+n(n-1)\mc{D}_{\a(2)}\mc{D}^{\b(2)}h_{\a(n-2)\b(2)}~, \label{ID0}
\end{align}
for any integer $n\geq 0$. It should be pointed out that the divergence in the second term of \eqref{ID0} is not defined for $n=0 $ and 1, however the corresponding numerical coefficient is equal to zero in both cases. 


\subsection{On-shell massive and partially massless fields} \label{section 2.2}

We define an on-shell (real) field $h_{\a(n)}$, with $n\geq 2 $, to be one which satisfies the first-order constraints\footnote{For $n=0$ only \eqref{OS1c} is present, whilst for $n=1$ there is only \eqref{OS1b}.}
\begin{subequations}\label{OS1}
\begin{align}
0&=\mc{D}^{\b\g}h_{\a(n-2)\b\g}~,\label{OS1a}\\
0&=\big(\mc{F}-\rho\big)h_{\a(n)} \label{OS1b}~, 
\end{align}
\end{subequations}
for some mass parameter  
$\rho \in \mb{R}$.
An equivalent form of the second irreducibility condition \eqref{OS1b} is
\begin{align}
\mc{D}_{(\a_1}{}^{\b}h_{\a_2\dots\a_n)\b}=\frac{\rho}{n} h_{\a(n)}~.
\label{OS1b+}
\end{align}
The on-shell field $h_{\a(n)}$ is said to be transverse with pseudo-mass $\rho$, spin $n/2$ and helicity $\s n/2$, where $\s=\rho/|\rho|$. 
The equations \eqref{OS1a} and \eqref{OS1b+} were introduced in 
\cite{BHRST} (see also \cite{BPSS}).\footnote{In the flat-space limit, these equations 
reduce to those considered in \cite{GKL,TV}.} 

From the relation \eqref{ID0} we see that when restricted to the space of transverse fields, the quadratic Casimirs \eqref{F} and \eqref{Q} are related via
\begin{align}
\mc{F}^2h_{\a(n)}=n^2\bigg(\mc{Q}-(n-2)(n+2)\mc{S}^2\bigg)h_{\a(n)}~.
\end{align}
Applying $\mc{F}$ to \eqref{OS1b}, we see that the on-shell conditions \eqref{OS1} lead to the following second-order mass-shell equation
\begin{align}
0=\bigg(\mc{Q}-\big[\big(\rho/n\big)^2+(n-2)(n+2)\mc{S}^2\big]\bigg)h_{\a(n)}~.\label{OS1c}
\end{align}
In terms of the AdS d'Alembertian, $\Box$, this reads
\begin{align}
0=\bigg(\Box - \big[(\rho/n)^2-2(n+2)\mc{S}^2\big]\bigg)h_{\a(n)}~.\label{OS1d}
\end{align}

An on-shell field $h^{(t)}_{\a(n)}$ is said to be partially massless\footnote{Partially massless fields have been studied in diverse dimensions for over 35 years, see \cite{DW2, DeserN1, DeserW1, DeserW4, Higuchi1, Higuchi2, Metsaev2, DNW} for some of the earlier works.
 Lagrangian models for partially massless fields in AdS$_d$ were constructed in \cite{Zinoviev, Metsaev, SV}, and in \cite{BSZ1,BSZ2} 
 for the specific case of AdS$_3$.  } 
with depth $t$ if, in addition to the conditions \eqref{OS1}, the pseudo-mass satisfies
\begin{align}
\rho  \equiv \rho^{(\pm)}_{(t,n)} = \pm n(n-2t)\mc{S}~,\qquad 1\leq t \leq \lfloor  n/2 \rfloor ~. \label{PM1}
\end{align}
We will refer to $\rho^{(\pm)}_{(t,n)}$ as the depth-$t$ pseudo-mass values.\footnote{We note that the $\pm$ in \eqref{PM1} coincides the sign of the helicity, since $\s= \rho^{(\pm)}_{(t,n)}/ |\rho^{(\pm)}_{(t,n)}|=\pm1$. The positive and negative branches are related via  the identity $\rho^{(+)}_{(\lfloor n/2 \rfloor+t,n)}=\rho^{(-)}_{(\lceil n/2 \rceil-t,n)}$ for arbitrary $t$.} The corresponding mass-shell equation \eqref{OS1c} reduces to 
\begin{align}
0=\big(\mc{Q}-\tau_{(t,n)}\mc{S}^2\big)h^{(t)}_{\a(n)} \label{PMWE}
\end{align}
where the constants $\tau_{(t,n)}$ are the partially massless values defined by
\begin{align}
\tau_{(t,n)}= 2n(n-2t) + 4 (t-1)(t+1)~. \label{PM2}
\end{align}
The following identity holds for any transverse field $h_{\a(n)}$
\begin{align}
\big(\mc{F}-\rho^{(+)}_{(t,n)}\big)\big(\mc{F}-\rho^{(-)}_{(t,n)}\big)h_{\a(n)}=n^2\big(\mc{Q}-\tau_{(t,n)}\mc{S}^2\big)h_{\a(n)}~, \label{ID6}
\end{align}
and can be used to factorise second order operators involving the partially massless values into first order ones.

Using the formalism developed in appendix \ref{appendix A}, it may be shown that at the partially massless points \eqref{PM1}, the system of equations \eqref{OS1} admits a depth-$t$ gauge symmetry of the form
\begin{align}
\delta_{\xi}h^{(t)}_{\a(n)}=\mc{D}_{(\a_1\a_2}\cdots\mc{D}_{\a_{2t-1}\a_{2t}}\xi_{\a_{2t+1}\dots\a_n)}~. \label{PMgt}
\end{align}
This is true only if the gauge parameter $\xi_{\a(n-2t)}$ is also on-shell with the same pseudo-mass\footnote{The system of equations \eqref{GPC1a} and \eqref{GPC1b} is defined only if $n-2t\geq 2$. For  $n-2t =1 $ only eq.  \eqref{GPC1b} is present. If $n-2t=0$, then these conditions must be replaced with $0=\big(\mc{Q}-(n-2)(n+2)\mc{S}^2\big)\xi$. }
\begin{subequations} \label{GPC1}
\begin{align}
0&= \mc{D}^{\b\g}\xi_{\a(n-2t-2)\b\g}~, \label{GPC1a}\\
0&=\big(\mc{F}-\rho^{(\pm)}_{(t,n)} \big)\xi_{\a(n-2t)}~. \label{GPC1b}
\end{align}
\end{subequations}
 We note that strictly massless fields $h_{\a(n)}$ correspond to those partially massless fields with the minimal depth $t=1$ and therefore have pseudo-mass equal to $\rho^{(\pm)}_{(1,n)}$. They are defined modulo the standard first order gauge transformations,
\begin{align}
\delta_{\xi}h_{\a(n)}=\mc{D}_{(\a_1\a_2}\xi_{\a_3\dots\a_n)}~.
\end{align} 

In the interest of making contact with the representation theory of the AdS$_3$ isometry group $\sSO(2,2)$, it is useful to recast eq. \eqref{OS1d} into the notation of \cite{BHRST}, 
\begin{align}
0=\bigg(\Box+ \ell^{-2}\big[n/2+1-\eta^{-2}\big]\bigg)h_{\a(n)}~,\qquad \ell^{-1}:= 2\mc{S}~,\qquad \eta^{-1}:=\frac{\rho }{2n\mc{S}}~.
\end{align}
In this context, the minimal energy of the corresponding $\mf{so}(2,2)$ irreducible representation\footnote{The unitary irreducible representations of $\mf{so}(2,2)$
are denoted $D(E_0,s)$, where $E_0$ is the minimal energy and $s$ the helicity, see \cite{DKSS} and references therein. The unitary bound is $E_0 \geq |s|$.}
 is $E_0=1+|\eta^{-1}|$, and the unitarity bound is $E_0\geq n/2$, or equivalently $|\rho|\geq n(n-2)\mc{S}$. The minimal energy of the  partially massless depth-$t$ field $h_{\a(n)}^{(t)}$ is $E^{(t,n)}_0=n/2+(1-t)$. Therefore, strictly massless fields saturate the unitarity bound, whilst true partially massless fields violate it (they are non-unitary). Finally, we note that the physical mass (see e.g. \cite{DeserW}) is related to the pseudo-mass via 
\begin{align}
\rho_{\text{phys}}^2=(\rho/n)^2-(\rho^{(\pm)}_{(1,n)}/n)^2~.
\end{align}
 In terms of $\rho_{\text{phys}}$, the unitarity  bound is $\rho_{\text{phys}}^2\geq 0$. Partially massless fields are thus seen to have negative physical mass-squared: $(\rho_{\text{phys}}^{(t,n)})^2=-4(t-1)(n-t-1)\mc{S}^2$. 

\subsection{Linearised higher-spin Cotton tensors}

It is a difficult technical problem to obtain an explicit expression for $\mf{C}_{\a(n)}(h)$ in terms of the AdS covariant derivative. In particular, the minimal uplift of \eqref{Mcot} is not gauge invariant or transverse. To restore these properties, \eqref{Mcot} must be supplemented with curvature dependent terms, as in the low spin cases \eqref{LowCot}.  However, for generic spin, previous attempts to do this were unsuccessful \cite{KP18,KP19}. In order to derive the Cotton tensors in AdS$_3$, it is advantageous to recast the problem into the framework of homogeneous polynomials (or generating functions). The basics of this formalism are discussed in appendix \ref{appendix A}. 

We begin by making an ansatz for the higher-spin Cotton tensor, which must be done separately for the bosonic ($n=2s$) and fermionic ($n=2s+1$) cases. In terms of homogeneous polynomials, our ans\"atze are as follows
\begin{subequations}\label{ansatz}
\begin{align}
\mf{C}_{(2s)}(h)&=\sum_{j=0}^{s-1}b_{j}\prod_{t=1}^{j}\bigg(\mc{Q}-\tau_{(s-t,2s)}\mc{S}^2\bigg) \mc{D}_{(2)}^{s-j-1}\mc{D}_{(0)}\mc{D}_{(-2)}^{s-j-1}h_{(2s)}~,\label{ansatzb}\\
\mf{C}_{(2s+1)}(h)&=\sum_{j=0}^{s}f_{j}\prod_{t=1}^{j}\bigg(\mc{Q}-\tau_{(s-t+1,2s+1)}\mc{S}^2\bigg) \mc{D}_{(2)}^{s-j}\mc{D}_{(-2)}^{s-j}h_{(2s+1)}~, \label{ansatzf}
\end{align}
\end{subequations}
for undetermined real coefficients $b_j$ and $f_j$, and where $\tau_{(t,n)}$ are the partially massless values \eqref{PM2}. We refer the reader to appendix \ref{appendix A} for an explanation of the notation. The main motivation for  
\eqref{ansatz}
is the expectation\footnote{Strictly speaking, to the best of our knowledge, most results regarding the factorisation of the conformal higher-spin kinetic operators apply only in even dimensions and only in the bosonic case.} that in an AdS background, the gauge invariant action \eqref{CSA} for a conformal higher-spin field $h_{\a(n)}$ should factorise (in the transverse gauge) into products of minimal second-order operators involving all partial mass values. The factorisation of the Cotton tensors will be elaborated on in the next section. 
 
 To fix the coefficients in \eqref{ansatz}, we require that the expressions \eqref{ansatzb} and \eqref{ansatzf} are transverse \eqref{cotT} and gauge invariant \eqref{cotG}. The former condition takes the form $0=\mc{D}_{(-2)}\mf{C}_{(n)}(h)$ which, upon employing the identity \eqref{ID15}, yields the recurrence relations
\begin{subequations}\label{ansatzT}
\begin{align}
0&=b_j-4(s-j)(s+j)b_{j-1}~,\qquad ~~~~\phantom{..}1\leq j \leq s-1~,\label{ansatzTb}\\
0&=f_j-4(s-j+1)(s+j)f_{j-1}~, \qquad 1\leq j\leq s ~.\label{ansatzTf}
\end{align}
\end{subequations}
 These two systems determine the coefficients $b_j$ and $f_j$ up to an overall normalisation:
 \begin{subequations}\label{coeffSol}
\begin{align}
b_j&=2^{2j}\binom{s+j}{2j+1}\frac{(2j+1)!}{s}b_0~,\\
f_j&=2^{2j}\binom{s+j}{2j}(2j)! f_0~.
\end{align} 
\end{subequations}
To ensure that $\mf{C}_{\a(n)}(h)$ reduces to \eqref{Mcot} in the flat limit, we choose $b_0$ and $f_0$ as follows
\begin{align}
b_0=\frac{1}{2^{2s-1}(2s-1)!}~,\qquad f_0=\frac{1}{2^{2s}(2s)!}~.
\end{align}
It is interesting to note that even though the imposition of transversality completely fixes the coefficients in \eqref{ansatz}, the resulting descendent is automatically gauge invariant (and vice versa). This can be attributed to the symmetry property \eqref{sym} and the Noether identity which follows as a consequence. Gauge invariance of \eqref{ansatz}, with the coefficients given by \eqref{coeffSol}, can also be checked explicitly by expressing the gauge transformations in the form $\delta_{\xi}h_{(n)} \propto\mc{D}_{(2)}\xi_{(n-2)}$ and using the identity \eqref{ID14}.
 
To conclude this section, we provide the 
final expressions for the higher-spin Cotton tensors  
with explicit indices.
They are given by  
\begin{subequations}\label{cotE}
\begin{align}
\mf{C}_{\a(2s)}(h)&=\frac{1}{2^{2s-1}}\sum_{j=0}^{s-1}2^{2j+1}\binom{s+j}{2j+1}\prod_{t=1}^{j}\bigg(\mc{Q}-\tau_{(s-t,2s)}\mc{S}^2\bigg) \non\\
&\phantom{\frac{1}{2^{2s-1}}\sum_{j=0}^{s-1}2^{2j+1}\binom{s+j}{2j+1}}\times\mc{D}_{\a(2)}^{s-j-1}\mc{D}_{\a}{}^{\b}\big(\mc{D}^{\b(2)}\big)^{s-j-1}h_{\a(2j+1)\b(2s-2j-1)}~, \label{cotF}\\
\mf{C}_{\a(2s+1)}(h)&=\frac{1}{2^{2s}}\sum_{j=0}^{s}2^{2j}\binom{s+j}{2j}\frac{(2s+1)}{(2j+1)}\prod_{t=1}^{j}\bigg(\mc{Q}-\tau_{(s-t+1,2s+1)}\mc{S}^2\bigg) ~~~~~~~~~~~~~~~~~~~~~~~~~~~\non\\
&\phantom{\frac{1}{2^{2s}}\sum_{j=0}^{s}\binom{s+j}{2j}\frac{(2s+1)}{(2j+1)}}\times\mc{D}_{\a(2)}^{s-j}\big(\mc{D}^{\b(2)}\big)^{s-j}h_{\a(2j+1)\b(2s-2j)}~, \label{cotB}
\end{align}
\end{subequations}
in the bosonic and fermionic cases respectively. 


\subsection{Factorisation of the Cotton tensors and CHS action} \label{factor1}

The higher-spin Cotton tensors \eqref{cotE} are both gauge invariant and transverse.
This means that the action for the conformal higher-spin field $h_{\a(n)}$,
\begin{align}
S_{\text{CS}}^{(n)}[h]=\frac{\text{i}^n}{2^{\lceil n/2 \rceil+1}}\int\text{d}^3x\, e \, h^{\a(n)}\mf{C}_{\a(n)}(h)~,\qquad e^{-1}:= \text{det}(e_a{}^{m})~,
\label{CSA}
\end{align}
which is of the Chern-Simons type, is gauge invariant. Since $h_{\a(n)}$ and $\mf{C}_{\a(n)}(h)$ are primary fields with dimension \eqref{1.7} and \eqref{1.9} respectively, 
the action is conformally invariant. 
Furthermore, \eqref{CSA} is symmetric in the sense that, modulo a total derivative, the relation
\begin{align} 
\int\text{d}^3x\, e \, g^{\a(n)}\mf{C}_{\a(n)}(h) = \int\text{d}^3x\, e \, h^{\a(n)}\mf{C}_{\a(n)}(g) \label{sym}
\end{align}
holds for arbitrary fields $g_{\a(n)}$ and $h_{\a(n)}$. 

Since the action is gauge invariant, we may impose the transverse gauge condition
\begin{align}
h_{\a(n)}\equiv h^{\text{T}}_{\a(n)}~,\qquad 0=\mc{D}^{\b(2)}h^{\text{T}}_{\b(2)\a(n-2)}~. \label{Tcond}
\end{align}
When the prepotential is transverse, only the $j=\lceil n/2 \rceil -1$ term contributes to \eqref{cotE}, and the corresponding Cotton tensors reduce to
\begin{subequations}\label{CF}
\begin{align}
\mf{C}_{\a(2s)}(h^{\text{T}})&=\prod_{t=1}^{ s-1 }\big(\mc{Q}-\tau_{(t,2s)}\mc{S}^2\big) \mc{D}_{\a}{}^{\b}h^{\text{T}}_{\a(2s-1)\b}~,\label{CFB}\\
\mf{C}_{\a(2s+1)}(h^{\text{T}})&=\prod_{t=1}^{ s}\big(\mc{Q}-\tau_{(t,2s+1)}\mc{S}^2\big) h^{\text{T}}_{\a(2s+1)}~. \label{CFF}
\end{align}
\end{subequations}
As a result, in this gauge the fermionic conformal higher-spin action \eqref{CSA} fully factorises into products of minimal second order operators involving partial mass values of all possible depths. From \eqref{CFB} it is clear that the bosonic Cotton tensor does not wholly factorise and does not include the maximal depth partial mass $\tau_{(s,2s)}$ as a factor. However, in this case one may show that the following descendent of $\mf{C}_{\a(2s)}(h)$ fully factorises
\begin{align}
 \mc{D}_{\a}{}^{\b}\mf{C}_{\a(2s-1)\b}(h^{\text{T}})=\prod_{t=1}^{ s }\big(\mc{Q}-\tau_{(t,2s)}\mc{S}^2\big) h^{\text{T}}_{\a(2s)}~,\label{FE}
\end{align}
and its factors include all possible partial mass values. 

Using the identity \eqref{ID6}, from \eqref{CF} one may derive the following useful alternative forms for the Cotton tensors,
\begin{subequations} \label{cotFO}
\begin{align}
\mf{C}_{\a(2s)}(h^{\text{T}})&=\frac{1}{(2s)^{2s-1}}\mc{F}\prod_{t=1}^{ s-1 }\big(\mc{F}-\rho^{(-)}_{(t,2s)}\big) \big(\mc{F}-\rho^{(+)}_{(t,2s)}\big)h^{\text{T}}_{\a(2s)}~, \label{cotFOB}\\
\mf{C}_{\a(2s+1)}(h^{\text{T}})&=\frac{1}{(2s+1)^{2s}}\prod_{t=1}^{ s }\big(\mc{F}-\rho^{(-)}_{(t,2s+1)}\big) \big(\mc{F}-\rho^{(+)}_{(t,2s+1)}\big)h^{\text{T}}_{\a(2s+1)}~. \label{cotFOF}
\end{align}
\end{subequations}
We see that $\mf{C}_{\a(n)}(h)$ factorises into products of first-order differential operators involving all partial pseudo-mass values.\footnote{We note that $\rho_{(s,2s)}^{(\pm)}=0$, so that the maximal-depth bosonic pseudo-mass also appears in \eqref{cotFOB}.} 


\subsection{Partially massless prepotentials}

From the above expressions for $\mf{C}_{\a(n)}(h)$, it is clear that the Cotton tensor of any on-shell partially massless field vanishes. More specifically, if $h_{\a(n)}\equiv h_{\a(n)}^{(t)}$ satisfies the on-shell conditions \eqref{OS1} and has depth-$t$, i.e. $\rho$ is given by \eqref{PM1}, then 
\begin{align}
\mf{C}_{\a(n)}(h)=0~.
\end{align}
In Minkowski space, it is known (see  \cite{HHL, HLLMP} and appendix \ref{AppendixC}) that the vanishing of the Cotton tensor associated with the field $h_{\a(n)}$, is a necessary and sufficient condition for $h_{\a(n)}$ to be pure gauge. As argued in appendix \ref{AppendixC}, an analogous statement holds in AdS space:
\begin{align}
\mf{C}_{\a(n)}(h)=0 \qquad \Longleftrightarrow \qquad h_{\a(n)}=\mc{D}_{\a(2)}\xi_{\a(n-2)}~, \label{PG}
\end{align}
 for some $\xi_{\a(n-2)}$. This provides another way to understand why partially massless fields in AdS$_3$ do not have any local dynamical degrees of freedom.

In fact, for on-shell partially massless fields $h_{\a(n)}^{(t)}$, the statement \eqref{PG} can be taken a step further: Writing  $h_{\a(n)}^{(t)}=\mc{D}_{\a(2)}\xi_{\a(n-2)}$ and requiring this to be transverse yields the equation
\begin{align}
0=\big(\tau_{(t,n)}-\tau_{(1,n)}\big)\mc{S}^2\xi_{\a(n-2)}-\frac{(n-2)(n-3)}{4(n-1)}\mc{D}_{\a(2)}\mc{D}^{\b(2)}\xi_{\a(n-4)\b(2)}~. \label{PGDT}
\end{align}
 Here we have made use of the 
 identities \eqref{ID14} and \eqref{PMWE}. Therefore, if $t=1$ (i.e. $h$ is massless) then $\xi_{\a(n-2)}$ is transverse. On the other hand, if $2\leq t \leq \lfloor n/2 \rfloor$ then \eqref{PGDT} implies
 \begin{align}
\xi_{\a(n-2)}=\mc{D}_{\a(2)}\xi_{\a(n-4)}\qquad \implies \qquad h_{\a(n)}^{(t)}=\mc{D}_{\a(2)}\mc{D}_{\a(2)}\xi_{\a(n-4)}~,
 \end{align}
 for some non-zero $\xi_{\a(n-4)}$. Once again, requiring $h_{\a(n)}^{(t)}$ to be transverse yields the equation
 \begin{align}
 0=\big(\tau_{(t,n)}-\tau_{(2,n)}\big)\mc{S}^2\mc{D}_{\a(2)}\xi_{\a(n-4)} - \frac{(n-4)(n-5)}{8(n-2)}\mc{D}_{\a(2)}\mc{D}_{\a(2)}\mc{D}^{\b(2)}\xi_{\a(n-6)\b(2)}~,
 \end{align} 
 and similar remarks hold. This procedure terminates when $t$ gradients have been extracted, whereupon $h_{\a(n)}^{(t)}$ takes the form
\begin{align}
h_{\a(n)}^{(t)}=\underbrace{\mc{D}_{\a(2)}\cdots\mc{D}_{\a(2)}}_{t-\text{times}}\xi_{\a(n-2t)}~,
\end{align}
 for some non-zero $\xi_{\a(n-2t)}$ being also on-shell (i.e. satisfying \eqref{GPC1}). The partially massless field $h_{\a(n)}^{(t)}$ will be said to be `pure depth-$t$ gauge'.


\section{Massive higher-spin gauge models} \label{section 3}

In this section we demonstrate that the properties of the higher-spin Cotton tensors allow us to analyse the on-shell dynamics of massive higher-spin gauge-invariant actions in AdS$_3$. 

\subsection{New topologically massive higher-spin gauge models}

The so-called `new topologically massive' (NTM) models for bosonic fields were first introduced in \cite{BKRTY} in Minkowski space. Extensions of these models to fields with half-integer spin were proposed in \cite{KP18}, where their generalisations to an AdS background were also given. 
These models are formulated solely in terms of the gauge prepotentials $h_{\a(n)}$ and 
the associated Cotton tensors $\mf{C}_{\a(n)}(h)$.  
Having obtained explicit expressions for the latter in AdS, we now take a closer look at these models.

Given an integer $n\geq 2$, the gauge-invariant NTM action for the field $h_{\a(n)}$ 
given in \cite{KP19} is  
\begin{align}
S_{\text{NTM}}^{(n)}[h]=\frac{\text{i}^n}{2^{\lceil n/2 \rceil+1}}\frac{1}{\rho}\int\text{d}^3x\, e \, \mf{C}^{\a(n)}(h) \big(\mc{F}-\rho\big)h_{\a(n)}~, \label{HSNTMG}
\end{align}
where $\rho$ is some arbitrary mass parameter. The equation of motion obtained by varying \eqref{HSNTMG} with respect to the field $h_{\a(n)}$ is 
\begin{align}
0=\big(\mc{F}-\rho\big)\mf{C}_{\a(n)}(h)~. \label{EOM1}
\end{align}
 We may analyse the solutions to \eqref{EOM1} in terms of (i) the gauge prepotential $h_{\a(n)}$; or (ii) its field strength  $\mf{C}_{\a(n)}(h)$.

Let us first work in terms of the prepotential. Since the action \eqref{HSNTMG} is invariant under the gauge transformations \eqref{GT}, we may impose the transverse gauge condition \eqref{Tcond}, whereupon the field equation \eqref{EOM1} takes the form
\begin{subequations}\label{EOM2}
 \begin{align}
 0&=\big(\mc{F}-\rho\big)\mc{F}\prod_{t=1}^{ s-1 }\big(\mc{F}-\rho^{(-)}_{(t,2s)}\big) \big(\mc{F}-\rho^{(+)}_{(t,2s)}\big)h^{\text{T}}_{\a(2s)}~,\label{EOM2a}\\ 
 0&=\big(\mc{F}-\rho\big)\prod_{t=1}^{ s }\big(\mc{F}-\rho^{(-)}_{(t,2s+1)}\big) \big(\mc{F}-\rho^{(+)}_{(t,2s+1)}\big)h^{\text{T}}_{\a(2s+1)}~, \label{EOM2b}
 \end{align}
 \end{subequations}
in the bosonic and fermionic cases respectively. For simplicity, here we will only seek particular solutions to \eqref{EOM2}, rather than general ones. Clearly, any $h_{\a(n)}^{\text{T}}$ satisfying
\begin{align}
0=\big(\mc{F}-\rho\big)h_{\a(n)}^{\text{T}}~,\qquad \rho\neq \rho_{(t,n)}^{(\pm)}~,\label{EOM3}
\end{align}
 is a solution to \eqref{EOM2}. In addition, any $h_{\a(n)}^{\text{T}}\equiv h_{\a(n)}^{(t)}$ satisfying 
 \begin{align}
0=\big(\mc{F}-\rho_{(t,n)}^{(\pm)}\big) h_{\a(n)}^{(t)}~,\label{EOM4}
\end{align}
 for some $1\leq t \leq \lfloor n/2 \rfloor$, is also a solution. However, in this case its Cotton tensor vanishes
\begin{align}
0=\mf{C}_{\a(n)}(h^{(t)})~. \label{HSCF}
\end{align}
As discussed section \ref{factor1}, this condition is necessary and sufficient to conclude that $h_{\a(n)}^{(t)}$ is pure depth-$t$ gauge. Hence, of these solutions, the only non-trivial one is \eqref{EOM3}, which propagates a single degree of freedom with pseudo-mass $\rho$, spin $n/2$ and helicity $ \text{sgn}(\rho) n/2$.

If instead we wish to analyse solutions to \eqref{EOM1} in terms of the field strength $\mf{C}_{\a(n)}(h)$, then the analysis should be split into two cases. First, if $\mf{C}_{\a(n)}(h)$ satisfies 
\begin{align}
0=\big(\mc{F}-\rho\big)\mf{C}_{\a(n)}(h)~, \qquad \rho\neq \rho_{(t,n)}^{(\pm)}~, \label{EOM58}
\end{align}
then, in accordance with section \ref{section 2.2}, eq. \eqref{EOM58} together with the conservation identity \eqref{cotT} means that the field strength $\mf{C}_{\a(n)}(h)$ itself defines an on-shell  spin $n/2$ field with pseudo-mass $\rho$, and helicity $ \text{sgn}(\rho) n/2$. Next we consider the case when $\mf{C}_{\a(n)}(h)$ satisfies 
\begin{align}
0=\big(\mc{F}-\rho_{(t,n)}^{(\pm)}\big)\mf{C}_{\a(n)}(h)~, \label{EOM59}
\end{align}
for some $1\leq t \leq \lfloor n/2 \rfloor$. 
 Now let us examine the Cotton tensor of the field strength $\mf{C}_{\a(n)}(h)$, which for ease of notation we denote $\mf{G}_{\a(n)}(h)$. Since $\mf{C}_{\a(n)}(h)$ is identically conserved, eq. \eqref{cotT}, its Cotton tensor factorises according to \eqref{cotFO}, 
\begin{subequations} \label{cotFOG}
\begin{align}
\mf{G}_{\a(2s)}(h)&=\frac{1}{(2s)^{2s-1}}\mc{F}\prod_{t=1}^{ s-1 }\big(\mc{F}-\rho^{(-)}_{(t,2s)}\big) \big(\mc{F}-\rho^{(+)}_{(t,2s)}\big)\mf{C}_{\a(2s)}(h)~, \label{cotFOGB}\\
\mf{G}_{\a(2s+1)}(h)&=\frac{1}{(2s+1)^{2s}}\prod_{t=1}^{ s }\big(\mc{F}-\rho^{(-)}_{(t,2s+1)}\big) \big(\mc{F}-\rho^{(+)}_{(t,2s+1)}\big)\mf{C}_{\a(2s+1)}(h)~. \label{cotFOGF}
\end{align}
\end{subequations}
On account of \eqref{EOM59}, we conclude that $\mf{G}_{\a(n)}(h)=0$. Therefore, $\mf{C}_{\a(n)}(h)$ is itself pure gauge: 
\begin{align}
\mf{C}_{\a(n)}(h)= \mc{D}_{\a(2)}\widehat{\mf{C}}_{\a(n-2)}(h)~,\label{EOM60}
\end{align}
 for some $\widehat{\mf{C}}_{\a(n-2)}(h)$. Since $\mf{C}_{\a(n)}(h)$ is gauge invariant, we can take $h_{\a(n)}$ in \eqref{EOM60} to be transverse, whereupon the left hand side of \eqref{EOM60} factorises. Consequently, the only way that $\mf{C}_{\a(n)}(h)$ can have the form \eqref{EOM60}, is if $h_{\a(n)}$ is pure gauge, and hence $\mf{C}_{\a(n)}(h)=0$. 
 
 Therefore, from both points of view, the conclusion is that if the pseudo-mass $\rho$ in \eqref{HSNTMG} takes on one of the partially massless values, then it describes only pure gauge degrees of freedom. If however $\rho$ satisfies $\rho\neq \rho_{(t,n)}^{(\pm)}$, but is otherwise arbitrary,\footnote{For unitarity one should restrict $\rho$ to satisfy $|\rho|> n(n-2)\mc{S}$.} then on-shell the model describes a spin $n/2$ mode with pseudo-mass $\rho$ and helicity $ \text{sgn}(\rho) n/2$.


\subsection{Topologically massive higher-spin gauge models}

Topologically massive models for higher-spin gauge fields are higher-spin extensions of linearised topologically massive gravity \cite{DJT1,DJT2}, and were proposed in \cite{KP18} in both Minkowski and AdS space. They are obtained by coupling the AdS$_3$ counterparts of 
the massless Fronsdal \cite{Fronsdal2} and  
Fang-Fronsdal \cite{FF2} actions to the bosonic (even $n$)  and fermionic 
(odd $n$) Chern-Simons action \eqref{CSA}, respectively. However, since the latter is constructed using the higher-spin Cotton tensor, whose explicit form was not known at the time of writing \cite{KP18}, the on-shell analysis was restricted to the case of Minkowski space only. In this section we address this issue and take a brief look at the on-shell content of these models in AdS$_3$.   

There are two types of the higher-spin massless actions, first-order and second-order ones. 
Given an integer $s\geq 2$, the first-order model is described by real fields 
$h_{\a(2s+1)},\,y_{\a(2s-1)}$ and $z_{\a(2s-3)}$, which are defined modulo gauge transformations of the form 
\begin{subequations} \label{FOGT}
\begin{align}
\d_{\xi} h_{\a(2s+1)}&=\mc{D}_{\a(2)}\xi_{\a(2s-1)} ~,\\
\d_{\xi} y_{\a(2s-1)}&=\frac{1}{(2s+1)(2s-1)}\big(\mc{F}-\rho^{(-)}_{(1,2s+1)}\big)\xi_{\a(2s-1)}~,\\
\d_{\xi} z_{\a(2s-3)}&=\mc{D}^{\b(2)}\xi_{\a(2s-3)\b(2)}~.
\end{align}
\end{subequations}
The three-dimensional counterpart of the  Fang-Fronsdal action is\footnote{There are actually two gauge invariant first-order Fang-Fronsdal actions in AdS$_3$ \cite{HHK}. The other is obtained by interchanging $\rho^{(+)}_{(1,2s+1)}$ with $\rho^{(-)}_{(1,2s+1)}$ everywhere in \eqref{FOGT} and \eqref{FOA}, or equivalently, sending $\mc{S}\rightarrow - \mc{S}$.}
\begin{align}
S_{\text{FF}}^{(2s+1)}[h,y,z]&=\bigg(-\frac{1}{2}\bigg)^s\frac{\ri}{2}\int \text{d}^3x\, e\, \bigg\{\frac{1}{(2s+1)}h^{\a(2s+1)}\big(\mc{F}-\rho^{(-)}_{(1,2s+1)}\big)h_{\a(2s+1)}  \non\\
&+4y^{\a(2s-1)}\big(\mc{F}-\rho^{(+)}_{(1,2s+1)}\big)y_{\a(2s-1)}-\frac{(s-1)}{s(2s-1)}z^{\a(2s-3)}\big(\mc{F}-\rho^{(-)}_{(1,2s+1)}\big)z_{\a(2s-3)} \non\\
&+2y^{\a(2s-1)}\bigg[(2s-1)\mc{D}^{\b(2)}h_{\a(2s-1)\b(2)}-\frac{(2s+1)(s-1)}{s}\mc{D}_{\a(2)}z_{\a(2s-3)}\bigg]\bigg\}~. \label{FOA}
\end{align}
This model can be reformulated in terms of a single reducible tensor field $\boldsymbol{h}_{\b(2),\a(n-2)}$, which corresponds to the AdS extension of the massless model introduced by Tyutin and Vasiliev \cite{TV}.

Given an integer $s\geq 2$, the second-order model is described by the real fields 
$h_{\a(2s)}$ and $y_{\a(2s-4)}$,  which are
defined modulo gauge transformations of the form 
\begin{subequations} \label{SOGT}
\begin{align}
\d_{\x} h_{\a(2s)}&=\mc{D}_{\a(2)}\xi_{\a(2s-2)}~, \label{SOGT1}\\
\d_{\x} y_{\a(2s-4)}&=\frac{2s-2}{2s-1}\mc{D}^{\b(2)}\xi_{\a(2s-4)\b(2)}~.
\end{align}
\end{subequations}
The corresponding gauge-invariant Fronsdal-type action is
\begin{align}
S_{\text{F}}^{(2s)}[h,y]&=\frac{1}{2}\bigg(-\frac{1}{2}\bigg)^s\int\text{d}^3x \, e \,\bigg\{h^{\a(2s)}\big(\mc{Q}-\tau_{(1,2s)}\mc{S}^2\big)h_{\a(2s)} +\frac{s}{2}h^{\a(2s)}\mc{D}_{\a(2)}\mc{D}^{\b(2)}h_{\a(2s-2)\b(2)} \non\\
&-\frac{(2s-3)}{2}y^{\a(2s-4)}\mc{D}^{\b(2)}\mc{D}^{\b(2)}h_{\a(2s-4)\b(4)}-\frac{(2s-3)}{s}\bigg[y^{\a(2s-4)}\big(\mc{Q}-\tau_{(2,2s)}\mc{S}^2\big)y_{\a(2s-4)}\non\\
&-\frac{(s-2)(2s-5)}{8(s-1)}y^{\a(2s-4)}\mc{D}_{\a(2)}\mc{D}^{\b(2)}y_{\a(2s-6)\b(2)}\bigg]\bigg\}~. \label{SOA}
\end{align}

For spin $s+1/2$ and $s$ respectively, the corresponding topologically massive models are described by the gauge-invariant actions
\begin{subequations}\label{TM}
\begin{align}
S_{\text{TM}}^{(2s+1)}[h,y,z]&= S^{(2s+1)}_{\text{CS}}[h]-\mu(\rho,s)S_{\text{FF}}^{(2s+1)}[h,y,z]~,\label{TMF}\\
S_{\text{TM}}^{(2s)}[h,y]&= S^{(2s)}_{\text{CS}}[h]-\nu(\rho,s)S_{\text{F}}^{(2s)}[h,y]~.\label{TMB}
\end{align}
\end{subequations}
The functions $\mu(\rho,s)$ and $\nu(\rho,s)$ are defined by 
\begin{subequations} \label{coupleC}
\begin{align}
\mu(\rho,s)&=\frac{1}{(2s+1)^{2s-1}}\big(\rho-\rho_{(1,2s+1)}^{(+)}\big)\prod_{t=2}^{s}\big[\rho^2-\big(\rho_{(t,2s+1)}^{(\pm)}\big)^2\big]~,\label{muF}\\
\nu(\rho,s)&=\frac{1}{(2s)^{2s-3}}\big(\rho-\rho_{(s,2s)}^{(\pm)}\big)\prod_{t=2}^{s-1}\big[\rho^2-\big(\rho_{(t,2s)}^{(\pm)}\big)^2\big]~,\label{muB}
\end{align}
\end{subequations}
where $\rho\in \mb{R}$ is some arbitrary real constant with mass dimension one. For unitarity, the mass parameter $\rho$ must satisfy $|\rho|> n(n-2)\mc{S}$. For such values the functions \eqref{muF} and \eqref{muB} are always non-vanishing. For the (non-unitary) true partially massless values $\rho=\rho_{(t,n)}^{(\pm)}$, with $2\leq t \leq \lfloor n/2 \rfloor$, the coupling constants \eqref{coupleC} vanish.

It may be shown that on-shell, the model $S_{\text{TM}}^{(n)}$ describes an irreducible spin-$n/2$ field with pseudo-mass $\rho$ and helicity whose sign is equal to $\rho/|\rho|$. Let us sketch how this can be seen in the bosonic case with $n=2s$. 

In analogy with the analysis of \cite{KP18},  using the equation of motion obtained by varying \eqref{TMB} with respect to $y_{\a(2s-4)}$, it follows that we may impose the gauge
\begin{align}
y_{\a(2s-4)}=0~,\qquad \mc{D}^{\b(2)}h_{\b(2)\a(2s-2)}=0~. \label{GF1}
\end{align} 
In this gauge the Cotton tensor factorises in accordance with section \ref{factor1}, and the equation of motion obtained by varying \eqref{TMB} with respect to $h_{\a(2s)}$ is
\begin{align}
0=\bigg(\mc{F}\prod_{t=2}^{s-1}\big(\mc{Q}-\tau_{(t,2s)}\mc{S}^2\big)+2s\nu(\rho,s)\bigg)\bigg(\mc{Q}-\tau_{(1,2s)}\mc{S}^2\bigg)h_{\a(2s)}~. \label{EOM5}
\end{align}
There are two types of solutions to \eqref{EOM5}. The first type solves $0=\big(\mc{Q}-\tau_{(1,2s)}\mc{S}^2\big)h_{\a(2s)}$, in which case $h_{\a(2s)}$ is massless with only pure gauge degrees of freedom. The second type of solutions are those satisfying
\begin{align}
0=\bigg(\mc{F}\prod_{t=2}^{s-1}\big(\mc{Q}-\tau_{(t,2s)}\mc{S}^2\big)+2s\nu(\rho,s)\bigg)h_{\a(2s)}~. \label{EOM6}
\end{align}
Once again, here we will only seek particular solutions to \eqref{EOM6}. For $\nu(\rho,s)$ defined as in \eqref{muB}, one can show that any $h_{\a(2s)}$ satisfying
\begin{align}
0=\big(\mc{F}-\rho\big)h_{\a(2s)} \quad \implies \quad 0=\big(\mc{Q}-[\rho^2/(2s)^2+4(s-1)(s+1)\mc{S}^2]\big)h_{\a(2s)}~,\label{sol1}
\end{align}
where $\rho$ satisfies $\rho \neq \rho_{(t,2s)}^{(\pm)}$, but is otherwise arbitrary, solves \eqref{EOM6}. We note that the residual gauge symmetry preserving the gauge fixing condition \eqref{GF1} is described by \eqref{SOGT1}, where the gauge parameter satisfies
\begin{align}
0=\mc{D}^{\b(2)}\xi_{\b(2)\a(2s-4)}~,\qquad 0=\big(\mc{F}-\rho_{(1,2s)}^{(\pm)}\big)\xi_{\a(2s-2)}~. \label{ResGS}
\end{align} 
For the massive solutions \eqref{sol1}, the residual gauge symmetry is exhausted since both \eqref{sol1} and \eqref{ResGS} imply $\xi_{\a(2s-2)}=0$ for $\rho \neq \rho_{(t,2s)}^{(\pm)}$.

 In accordance with section \ref{section 2.2}, the field $h_{\a(2s)}$ satisfying the second equation in \eqref{GF1} and the first equation in \eqref{sol1} defines an on-shell spin-$s$ field with pseudo-mass $\rho$ and helicity $s\rho/|\rho|$.  It is important to note that the above discussion assumes that the pseudo-mass does not take on any of the true partially massless values. If this is the case then the coupling constant  $\nu$ vanishes, and we are left with the Chern-Simons sector, which describes only pure gauge degrees of freedom.  The analysis in the fermionic case proceeds in a similar fashion and we do not repeat it here.


\section{Supersymmetric higher-spin gauge models in AdS$_3$} \label{section 4}

In this section we present a closed-form expression for the linearised higher-spin super-Cotton tensor in $\mc{N}=1$ AdS superspace,  AdS$^{3|2}$,  
and give several applications of this result. Before turning to the main construction, 
some general comments are in order about the super-Cotton tensor in supergravity.

\subsection{Conformal supergravity and super-Cotton tensors}

When formulating three-dimensional $\cN$-extended conformal supergravity in conformal superspace \cite{BKNT-M1}, all information 
about the corresponding  superspace geometry is encoded in a single curvature superfield, which is the super-Cotton 
tensor $\cW$. 
The functional structure of the super-Cotton tensor is $\cN$-dependent. 
 The $\cN=1$ super-Cotton tensor \cite{KT-M12}
 is a primary symmetric rank-3 spinor superfield
 $\cW_{\a\b\g}$ of dimension 5/2, which obeys the conformally invariant constraint
 \cite{BKNT-M1}
\bea
\nabla^\a \cW_{\a \b\g} = 0 ~. 
\eea
In the $\cN=2$ case,  the super-Cotton tensor \cite{ZP,Kuzenko12} 
is a primary symmetric rank-2 spinor superfield 
 $\cW_{\a\b}$ of dimension 2, which obeys the Bianchi identity  \cite{BKNT-M1}
\bea
\nabla^{\a I} \cW_{\a\b} = 0 ~. 
\eea
In the $\cN=3$ case,  the super-Cotton tensor is a  primary spinor superfield
 $\cW_{\a}$ of dimension 3/2 constrained by \cite{BKNT-M1}
 \bea
\nabla^{\a I} \cW_\a = 0 ~.
\eea
In the $\cN=4$ case, the super-Cotton tensor is a primary  scalar
 superfield  $\cW$ of dimension 1
constrained by \cite{BKNT-M1}
\bea
\nabla^{\a I}\nabla_{\a}^J \cW=\frac{1}{4}\d^{IJ}\nabla^{\a K}\nabla_{\a}^K \cW~.
\eea
For $\cN>4$,  the super-Cotton tensor \cite{HIPT,KLT-M11}
is a completely antisymmetric tensor $\cW^{IJKL}$
of dimension 1 constrained by \cite{BKNT-M1}
\bea \nabla_{\a}^I \cW^{JKLP} = \nabla_\a^{[I} \cW^{JKLP]} 
- \frac{4}{\cN - 3} \nabla^Q_{\a} \cW^{Q [JKL} \d^{P] I} \ .
\eea
In the above relations, $ \nabla^I_\a$ denotes the spinor covariant derivative
of $\cN$-extended conformal superspace \cite{BKNT-M1}.

The super-Cotton tensor is naturally encoded in the action for conformal 
supergravity,\footnote{For the $1\leq \cN \leq 6$ cases, 
the complete nonlinear actions for $\cN$-extended conformal supergravity were
derived in \cite{BKNT-M2,KNT-M14} using the off-shell formulation for 
$\cN$-extended conformal supergravity developed in \cite{BKNT-M1}.
The $\cN=1$ and $\cN=2$ conformal supergravity theories were constructed for the first time by  van Nieuwenhuizen \cite{vN}, and by
Ro\v{c}ek and  van Nieuwenhuizen  \cite{RvN}, respectively. 
The off-shell action for
$\cN=6$ conformal supergravity was independently derived by 
Nishimura and Tanii \cite{NT}.
On-shell  formulations for $\cN$-extended conformal supergravity with  $\cN>2$
were given in \cite{LR89,NG}.} 
$S_{\rm CSG}$,  in the following sense 
 \cite{BKNT-M1,KNT-M14}:
\bea
\cW \propto \frac{\d S_{\rm CSG}}{\d H}~.
\eea
Here $H$ denotes an unconstrained gauge prepotential for conformal supergravity. 
Modulo purely gauge degrees of freedom,
 the structure of unconstrained conformal gauge prepotentials are as follows:
 $H_{\a\b\g} $ 
for $\cN=1$ \cite{GGRS}, $H_{\a\b}$ for $\cN=2$ \cite{ZP,Kuzenko12}, 
$H_\a$ for $\cN=3$ \cite{BKNT-M1}, and $H$ for $\cN=4$ \cite{BKNT-M1}. 

In the case of $\cN$-extended Poincar\'e supersymmetry, higher-spin generalisations of the linearised super-Cotton tensor were given in 
\cite{K16,KT} for $\cN=1$, in \cite{KO} for $\mc{N}=2$ and in \cite{BHHK} for  $\cN>2$. 
Below we generalise the results of \cite{K16,KT} to the $\cN=1$ AdS supersymmetry, 
while the $\cN>1$ case will be discussed elsewhere. 


\subsection{Higher-spin super-Cotton tensors in conformally flat superspace}

In $\cN=1$ supergravity,  
the  covariant derivatives\footnote{In the hope that no confusion arises, we use the same notation for the vector derivative \eqref{VecDer} and its $\mc{N}=1$ analogue. }
\bea
\mc{D}_A = (\mc{D}_a, \mc{D}_\a) = E_A  + \O_A 
=E_A{}^M {\partial}_M + \hf \O_A{}^{bc}M_{bc}~, \label{N10}
\eea
obey the following graded commutation relations \cite{GGRS,KLT-M11}
\bsubeq \label{N=1alg}
\bea
\{\cD_\a,\cD_\b\}&=&
2\ri\cD_{\a\b}
-4\ri \cS M_{\a\b}
~,~~~~~~~~~
\label{N=1alg-a}
\\
\left[ \cD_{\a \b} , \cD_\g \right] &=& 
- 2 \ve_{\g(\a} \cS \cD_{\b)} + 2 \ve_{\g(\a} \cC_{\b) \d \r} M^{\d\r}
\non\\
&&
+ \frac{2}{3} \big( \cD_\g \cS M_{\a\b} 
- 4 \cD_{(\a} \cS M_{\b) \g} \big) 
~,~~~~~~~~~
\label{N=1alg-b}
\eea
\esubeq
Here the torsion superfields $\cS$ and  $\cC_{\a\b\g}=\cC_{(\a\b\g)}$ 
are real and related to each other by  the Bianchi identity
$
\cD^\g \cC_{\a\b\g} = -\frac{4\ri }{3} \cD_{\a\b} \cS $.

The algebra of covariant derivatives \eqref{N=1alg} is invariant under 
the following super-Weyl transformation \cite{ZP,ZP89,LR-brane}
\bsubeq \label{2.10}
\bea
\d_\S\cD_\a&=&
\hf \S\cD_\a + \cD^{\b}\S M_{\a\b}
~,
\\
\d_\S\cD_a&=&
\S\cD_a
+\frac{\ri}{ 2}(\g_a)^{\g\d}\cD_{\g} \S\cD_{\d}
+\ve_{abc}\cD^b\S M^{c}
~,
\eea
\esubeq
where the parameter $\S$  is a real unconstrained superfield.
 The corresponding super-Weyl transformations of the torsion tensors are
\bea
\d_\S\cS&=&\S\cS-\frac{\ri}{4}  \cD^2\S~,~~~~~~
\d_\S \cC_{\a\b\g}=\frac{3}{2}\S \cC_{\a\b\g}-\frac{\ri}{2}  \cD_{(\a\b}\cD_{\g)}\S
~.
\eea
The $\cN=1$ super-Cotton tensor  \cite{KT-M12} is given by the expression
 \bea
\cW_{\a\b\g} = \Big(\frac{\ri }{2} \cD^2 +4\cS\Big) \cC_{\a\b \g} 
+  \ri \cD_{(\a\b} \cD_{\g)} \cS 
~.
\eea
Its super-Weyl transformation  is
$\d_\S \cW_{\a\b\g} = \frac{5}{2} \S \cW_{\a\b\g}$, which means that $\cW_{\a\b\g} $
is a primary superfield.

Let $\cM^{3|2}$ be a conformally flat superspace, 
\bea
\cW_{\a\b\g} =0~.
\eea
A real supervector field $\z= \z^B E_B$  is called conformal Killing if 
\begin{subequations} \label{CCSVF}
\bea
\Big[ \z^B \cD_B + \hf K^{\b \g} [\z] M_{\b\g} , \cD_A\Big]  + \delta_{\S [\z]}  \cD_{A} = 0 ~,
 \label{CCSVF.a}
\eea
for some Lorentz ($K^{\b \g}[\z]$) and super-Weyl ($\S[ \z] $) parameters. 
This equation implies that $\z^{\b \g} $ is the only independent transformation parameter, 
\bea
\z^\b = \frac{\ri}{6} \cD_\g \z^{\b \g} ~,\qquad K_{\b\g} [\z] = 2 \cD_{(\b} \z_{\g) }-2\mc{S}\z_{\b\g} ~,
\qquad \S[\z] = \cD_\b \z^\b = \frac 13 \cD_b \z^b~,
 \label{CCSVF.b}
\eea
and it obeys the superconformal Killing equation
\bea
\cD_{(\a} \z_{\b\g ) } =0 \quad \implies \quad \cD_{(\a\b} \z_{\g \d) } =0 
~~ \Longleftrightarrow ~~  \cD^a \z^b + \cD^b \z^a = 2 \eta^{ab} \S[\z] ~.
 \label{CCSVF.c}
\eea
\end{subequations}
The conformal Killing supervector fields of $\cM^{3|2}$ span the conformal superalgebra of $\cM^{3|2}$, 
which is isomorphic to  $\mathfrak{osp}(1|4, {\mathbb R})$.
 A primary tensor superfield $\F$ (with suppressed indices) possesses the following superconformal transformation law
\bea
-\d_\z \F = \Big(\z^B \cD_B +\hf K^{\b\g}[\z] M_{\b\g} + d_\F \S[\z]\Big) \F ~, 
\label{414}
\eea
where $d_\F$ is the dimension of $\F$. It should be pointed out that the conformal Killing supervector field $\z^B$ contains two independent component fields, $\z^b |$ and $\z^\b|$.\footnote{As usual, the bar-projection of a tensor superfield ${V} ={V}(x,\q)$ (with suppressed indices) is defined by 
${V}|:= { V}(x,\q )\big|_{\q =0}$.}

Let us fix a positive integer $n\geq 1$. The conformal 
gauge supermultiplet
 ${H}_{\a(n) } $ is a real primary superfield of dimension
\bea
d_{H_{\a(n)}} = 1-\frac{n}{2}  
\label{4.17}
\eea
with the gauge transformation law
\bea
\d_\L {H}_{\a(n) } =\ri^n \cD_{(\a_1} \L_{\a_2 \dots \a_n) }~,
\label{sgt}
\eea
with the  gauge parameter $\L_{\a(n-1)}$
being real but otherwise unconstrained.
In accordance with \cite{KP18}, the super-Cotton tensor $\mf{W}_{\a(n)}(H)$ associated with  $H_{\a(n)}$ 
is defined to satisfy the following three conditions: 
(i) it is a primary descendent of $H_{\a(n)}$; (ii) it is transverse (or covariantly conserved),
\begin{align}
0=\mc{D}^{\b}\mf{W}_{\b\a(n-1)}(H) ~;
\end{align}
(iii) it is invariant under the gauge transformations
 \eqref{sgt},
\begin{align}
0=\mf{W}_{\a(n)}(\delta_{\L }H)~. 
\end{align}
The conditions (i) and (ii)    
fix the dimension of $\mf{W}_{\a(n)}(H)$
to be
\bea
d_{\mf{W}_{\a(n)}} = 1 +\frac{n}{2} ~.
\label{4.21}
\eea
It was argued in \cite{KP18} that the supermultiplets $\mf{W}_{\a(n)}(H)$ exist for $n\geq 1$ in any conformally flat superspace $\cM^{3|2}$. In terms of conformally covariant derivatives $\nabla_A = (\nabla_a, \nabla_\a)$ \cite{BKNT-M1}, the super-Cotton tensors were constructed in \cite{KP19}. Below we will derive closed-form expressions for $\mf{W}_{\a(n)}(H)$ in terms of the Lorentz covariant derivatives 
$\cD_A$ for the case $\cM^{3|2}={\rm AdS}^{3|2}$, i.e. the  $\cN=1$ AdS superspace. 


\subsection{$\mc{N}=1$ AdS superspace geometry}

 The geometry of 
 AdS$^{3|2}$ 
 is encoded in its covariant derivatives, $\cD_A$,
with the graded commutation relations \cite{GGRS,KLT-M12}
\begin{subequations} \label{algN1}
\begin{align}
\{ \mc{D}_\a , \mc{D}_\b \} &= 2\ri \mc{D}_{\a\b} - 4\ri\mc{S} M_{\a\b}~, \\
\ [ \mc{D}_{\a \b}, \mc{D}_\g ] &= -2\mc{S} \ve_{\g(\a}\mc{D}_{\b)}~, \\
\ [ \mc{D}_{\a \b}, \mc{D}_{\g \d} ] &= 4 \mc{S}^2 \Big(\ve_{\g(\a}M_{\b)\d} + \ve_{\d(\a} M_{\b)\g}\Big) \label{algN1c}
~,
\end{align}
\end{subequations}
with $\cS \neq 0$ a constant real parameter, 
which 
may be positive or negative. The two choices, $\cS=|\cS|$ and $\cS = -|\cS|$,  correspond to the so-called $(1,0)$ and $(0,1)$  
AdS superspaces \cite{KLT-M12}, which are different realisations of $\cN=1$ AdS superspace. The $(1,0)$ and $(0,1)$ AdS superspaces are naturally embedded 
in $(1,1)$ AdS superspace \cite{KLT-M12,HHK} and are related to each other by a parity transformation.\footnote{The only difference between the $(1,0)$ and $(0,1)$ AdS supersymmetry types is the fact that the mass terms in the corresponding Killing spinor equations \eqref{440} have different signs.}

Isometries of AdS${}^{3|2}$ are generated by Killing supervector fields $\z^B$
on AdS${}^{3|2}$. They are defined to be
those  conformal Killing supervector fields, eq. \eqref{CCSVF}, which obey the additional restriction $\S[\z]=0$, 
\bea
\Big[ \z^B \cD_B + \hf K^{\b \g} [\z] M_{\b\g} , \cD_A\Big]   = 0 ~.
 \label{421}
\eea 
The important properties of the Killing supervector fields include the following:
\begin{subequations}\label{4.21abc}
\bea
\cD_\a \z_{\b\g} &=& 2\ri ( \ve_{\a\b} \z_\g + \ve_{\a\g} \z_\b) ~,\\
\cD_\a \z_\b &=& \hf K_{\a\b} [\z]   + \cS \z_{\a\b}~,\\
\cD_\a K_{\b\g} [\z]&=& 4\ri \cS( \ve_{\a\b} \z_\g + \ve_{\a\g} \z_\b) ~,
\eea
\end{subequations}
and their corollary
\bea
\big( \ri \cD^2 + 12 \cS \big) \z_\a=0~,
\eea
where we have denoted $\mc{D}^2 = \mc{D}^\a \mc{D}_\a$. 
The relations \eqref{4.21abc} tell us that the $\q$-dependence of the Killing superfield
parameters  $\z^B(x,\q)$ and $K^{\b\g}(x,\q) $  in \eqref{421} is determined by their values at $\q=0$. 
 Given a tensor superfield $\F$ on AdS${}^{3|2}$, 
its AdS transformation law is obtained from \eqref{414} by setting $\S[\z]=0$, 
\bea
-\d_\z \F = \Big(\z^B \cD_B +\hf K^{\b\g}[\z] M_{\b\g} \Big) \F ~.
\label{424}
\eea
We recall that the independent component fields of $\z^B$ are 
$\z^b |$ and $\z^\b|$.  These components are, respectively,  a Killing vector field and a Killing spinor field on AdS$_3$. The equation on the Killing spinor field follows from projecting
\bea
\cD_{\a\b} \z_\g = - \cS (\ve_{\g \a} \z_\b + \ve_{\g\b} \z_\a) \quad 
\Longleftrightarrow \quad \cD_a \z_\b = \cS (\g_a)_\b{}^\g \z_\g~, 
\eea
which is a simple corollary of \eqref{4.21abc}.

There are two independent quadratic Casimir operators\footnote{For the symplectic groups our notation is $\sSp (2n, {\mathbb R})  \subset \sGL (2n,{\mathbb R})$, 
hence  $\sSp (2, {\mathbb R} ) \cong  \sSL (2,{\mathbb R} ) \cong \sSU(1,1)$.}
 of the $\cN=1$ AdS isometry supergroup, $\sOSp (1|2; {\mathbb R} ) \times  \sSL (2,{\mathbb R} ) $,
 which may be chosen as follows:
\begin{subequations}
\begin{align}
\mathbb{F}&:=-\frac{\rm{i}}{2}\mc{D}^2+2\mc{D}^{\a\b}M_{\a\b}~,\qquad \big[\mathbb{F},\mc{D}_A\big]=0~, \label{SF}\\
\mathbb{Q}&:=-\frac{1}{4}\mc{D}^2\mc{D}^2+\rm{i}\mc{S}\mc{D}^2~,\qquad ~~\big[\mathbb{Q},\mc{D}_A\big]=0~. \label{SQ}
\end{align}
\end{subequations}
The second of these may be expressed in terms of the AdS d'Alembertian, $\Box = \cD^a \cD_a$, via the relation
\begin{align}
-\frac{1}{4} \cD^2 \cD^2
=  \Box -2\ri {\cS} \cD^2 
+2 {\cS} \cD^{\a \b} M_{\a\b} -2 {\cS}^2 M^{\a\b} M_{\a\b}~.
\end{align}
The fact that they are Casimir operators follows from  
the useful identities:
\begin{subequations}  \label{A8-mod}
	\bea 
	\mc{D}_\a \mc{D}_\b &=& \ri \mc{D}_{\a\b} - 2\ri\mc{S}M_{\a\b}+\frac{1}{2}\ve_{\a\b}\mc{D}^2~, \\
	\mc{D}^{\b} \mc{D}_\a \mc{D}_\b &=& 4\ri \mc{S}\mc{D}_\a~, \quad \{ \mc{D}^2, \mc{D}_\a  \} = 4\ri \mc{S}\mc{D}_\a~,\\
 \mc{D}^2 \mc{D}_\a &=& 2\ri \mc{S}\mc{D}_\a + 2\ri \mc{D}_{\a\b} \mc{D}^\b - 4\ri \mc{S} \mc{D}^\b M_{\a\b}~, \\
\qquad \ [ \mc{D}_{\a} \mc{D}_{\b}, \mc{D}^2 ]
&=& 0 \quad \Longrightarrow \quad \ [\mc{D}_{\a\b}, \mc{D}^2 ] = 0~,\\
\ [ \mc{D}_\a ,  \Box] &=& 2\mc{S} \mc{D}_{\a\b} \mc{D}^\b + 3\mc{S}^2 \mc{D}_\a~.
	\eea
\end{subequations}
It is an instructive exercise to check these identities. 

\subsection{On-shell massive and partially massless superfields in AdS$_3$}

We define an on-shell real superfield $H_{\a(n)}$, with $n\geq 1 $, to be one which satisfies the constraints
\begin{subequations}\label{OS2}
\begin{align}
0&=\mc{D}^{\b}H_{\a(n-1)\b}~,\label{OS2a}\\
0&=\big(\mb{F}- M\big)H_{\a(n)} \label{OS2b}~, 
\end{align}
\end{subequations}
for some mass parameter $M$ which can take any real value, $M \in \mb{R}$. The field $H_{\a(n)}$ is said to be transverse with pseudo-mass $M$, superspin $n/2$ and superhelicity $\frac{1}{2}\big(n+\frac{1}{2}\big)\s$, where $\s=M/|M|$. It can be shown that any superfield satisfying \eqref{OS2a} also satisfies
\begin{align}
-\frac{\rm{i}}{2}\mc{D}^2H_{\a(n)}=\bigg(\mc{D}_{(\a_1}{}^{\b}+(n+2)\mc{S}\delta_{(\a_1}{}^{\b}\bigg)H_{\a_2\dots\a_n)\b}~. \label{Dsquare}
\end{align}
Furthermore, if it satisfies both \eqref{OS2a} and  \eqref{OS2b}, then as a consequence we have
\begin{align}
-\frac{\rm{i}}{2}\mc{D}^2H_{\a(n)}&=\frac{1}{2n+1}\bigg( M+2n(n+2)\mc{S}\bigg)H_{\a(n)}~.\label{OS2c}
\end{align}
From this we can deduce that the second-order wave equation which $H_{\a(n)}$ satisfies is
\begin{align}
0=\bigg(\mb{Q}-\frac{1}{(2n+1)^2}\big[ M+2n(n+2)\mc{S}\big]\big[M+2(n-1)(n+1)\mc{S}\big]\bigg)H_{\a(n)}~. \label{OS2d}
\end{align}
The equations \eqref{OS2a} and \eqref{OS2c} were introduced in \cite{KNT-M}.
In the flat superspace limit, they reduce to the mass-shell equations given in \cite{KT}.

A novel feature of three-dimensional field theories in AdS superspace is the existence of two distinct types of partially massless superfields. We will say that an on-shell superfield has type A or type B partially massless symmetry if, in addition to \eqref{OS2}, its pseudo mass satisfies 
\begin{subequations} \label{SPM}
\begin{align}
M\equiv M^{(A)}_{(t,n)}&=-2\big[n(n-2t)-(t+1)\big]\mc{S}~,\qquad ~~~~~ 0 \leq t \leq \lceil n/2 \rceil -1~,\label{SPMA}\\
M\equiv M^{(B)}_{(t,n)}&=\phantom{-}2\big[n(n-2t+1)-(t-1)\big]\mc{S}~,\qquad 1\leq t \leq \lfloor n/2 \rfloor~, \label{SPMB}
\end{align}
\end{subequations}
respectively. We will refer to these as type A and type B pseudo masses, and we say that the corresponding superfield carries super-depth $t$. The latter is an integer whose range of allowed values for each type is specified  in \eqref{SPM}. The relation
\begin{align}
M^{(A)}_{(\lceil n/2 \rceil +t-1,n)}=M^{(B)}_{(\lfloor n/2 \rfloor -t+1,n)} \label{ID9}
\end{align}
 holds between the  two types of masses for all $t$. The second order wave equation \eqref{OS2d} satisfied by partially massless superfields reduces to
\begin{align}
0=\big(\mb{Q}-\lambda^{(A)}_{(t,n)}\mc{S}^2\big)H^{(t,A)}_{\a(n)}~,\qquad 0=\big(\mb{Q}-\lambda^{(B)}_{(t,n)}\mc{S}^2\big)H^{(t,B)}_{\a(n)}~,
\end{align}
where the constants 
\begin{align}
\lambda^{(A)}_{(t,n)}=4t(t+1)~, \qquad \lambda^{(B)}_{(t,n)}=4(n-t)(n-t+1)~,
\end{align}
will be referred to as the type A and type B partially massless values.

It may be shown that the system of equations satisfied by on-shell type A and type B partially massless superfields, $H^{(t,A)}_{\a(n)}$ and $H^{(t,B)}_{\a(n)}$ respectively, admit the gauge symmetries
\begin{subequations} \label{SPMG}
\begin{align}
\delta_{\L}H^{(t,A)}_{\a(n)}&=\text{i}^n\mc{D}_{(\a_1\a_2}\cdots\mc{D}_{\a_{2t-1}\a_{2t}}\mc{D}_{\a_{2t+1}}\L_{\a_{2t+2}\dots\a_n)}~,\label{SPMG1}\\
\delta_{\L}H^{(t,B)}_{\a(n)}&=\phantom{\rm{i}^n}\mc{D}_{(\a_1\a_2}\cdots\mc{D}_{\a_{2t-1}\a_{2t}}\L_{\a_{2t+1}\dots\a_n)}~.\label{SPMG2}
\end{align}
\end{subequations}
Similar to the non-supersymmetric case (see eq. \eqref{GPC1}), this is true only if the gauge parameters $\L_{\a(n-2t-1)}$ and $\L_{\a(n-2t)}$ are also on-shell with the same pseudo-mass as their gauge field. We point out that strictly massless superfields $H_{\a(n)}$, defined modulo the standard gauge transformations \eqref{sgt},
correspond to type A partially massless superfields with the minimal super-depth of $t=0$.


\subsection{Component analysis}

In order to make the content of the two types of partially massless supermultiplets more transparent, in this section we study their component structure in more detail. 

We start with some general comments.
Every Killing supervector field $\z^B$ on AdS$^{3|2}$ can be uniquely decomposed as a sum 
\bea
\z^B= \z^B_{(\text{even})}+\z^B_{(\text{odd})}~,
\eea
 where $\z^B_{(\text{even})}$ and $\z^B_{(\text{odd})}$ are 
  even $\big( v^b:=\z^b_{(\text{even})}| \neq 0 $ and $ \z_{(\text{even})}^\b| = 0 \big)$ and odd 
 $\big( \z_{(\text{odd})}^b| = 0 $ and $ \e^{\b}:=\z_{(\text{even})}^\b| \neq 0\big)$
 Killing supervector fields, respectively. 
 Here $v^b(x) $ is a Killing vector field on AdS$_3$, 
\bea
 \cD^a v^b + \cD^b v^a = 0~,
 \eea
 and $\e^\b(x)$ is a Killing spinor field on AdS$_3$,
 \bea
\cD_{\a\b} \e_\g = - \cS (\ve_{\g \a} \e_\b + \ve_{\g\b} \e_\a) ~.
\label{440}
\eea

Let us consider a tensor superfield $\F$ on AdS${}^{3|2}$ with  
the transformation law  \eqref{424}.
Its independent component fields are contained in the set of fields 
$\vf=\mathfrak{V}|$, where  
$\mathfrak{V}:=\big\{ \F, \cD_{ \a} \F, \cdots\big\}$.
Choosing $\z^B$ in \eqref{424} to be $ \z^B_{(\text{even})}$, one observes that the component fields $\vf$ transform as tensor fields on AdS$_3$,
\bea
-\d_v \vf = \Big(v^b \cD_b +\hf K^{\b\g}[v] M_{\b\g} \Big) \vf ~. 
\eea
Choosing $\z^B$ in \eqref{424} to be $ \z^B_{(\text{odd})}$, we find that 
the supersymmetry transformation laws of the component fields
are
given by\footnote{To derive this transformation rule, the relation $K_{\a\b}[\z]=\frac{1}{2}\mc{D}^{\g}{}_{(\a}\z_{\b)\g}$ may be useful. }
\bea
-\d_\e \vf =  \e^\b (\cD_\b \mathfrak{V})| ~.
\label{susyT}
\eea

If $H_{\a(n)}$ is an on-shell superfield on AdS$^{3|2}$ with pseudo-mass $M$, then it has only two independent component fields, which we define according to 
\begin{align}
h_{\a(n)}:=H_{\a(n)}|~,\qquad \psi_{\a(n+1)}:=\text{i}^{n+1}\mc{D}_{(\a_1}H_{\a_2\dots\a_{n+1})}| 
=\text{i}^{n+1}\mc{D}_{\a_1}H_{\a_2\dots\a_{n+1}}| 
~.
\label{443}
\end{align} 
As a consequence of \eqref{OS2a} and the identities \eqref{A8-mod}, both component fields are transverse
\begin{align}
0=\mc{D}^{\b\g}h_{\b\g\a(n-2)}~,\qquad 0=\mc{D}^{\b\g}\psi_{\b\g\a(n-1)}~. \label{OS3a}
\end{align}
To deduce the other first order constraint which they satisfy, the relation 
\begin{align}
\mb{F}\Phi_{\a(n)}|=\bigg(\frac{2n+1}{n}\mc{F}+(n+2)\mc{S}\bigg)\Phi_{\a(n)}|~, \label{ID1}
\end{align}
which holds for an arbitrary transverse superfield $\Phi_{\a(n)}$, is useful. Making use of \eqref{ID1}, we find that the component fields satisfy
\begin{subequations}\label{OS3b}
\begin{align}
0&=\bigg(\mc{F}-\frac{n}{2n+1}\big[M-(n+2)\mc{S}\big]\bigg)h_{\a(n)}~,\\
0&=\bigg(\mc{F}-\frac{n+1}{2n+1}\big[M+(n-1)\mc{S}\big]\bigg)\psi_{\a(n+1)}~.
\end{align}
\end{subequations}
The equations \eqref{OS3a} and \eqref{OS3b} define on-shell fields in accordance with section \ref{section 2.2}. Finally, making use of \eqref{susyT} and \eqref{OS2}, the supersymmetry transformation laws of the component fields
 prove to be
\begin{subequations}
\begin{align}
-\delta_{\e}h_{\a(n)}&= (-\ri)^{n+1}\e^{\b}\psi_{\b\a(n)}~,\\
-\delta_{\e}\psi_{\a(n+1)}&=-\ri^n\e^{\b}\mc{D}_{\b\a}h_{\a(n)}+\frac{\ri^n}{(2n+1)}\Big(M+n(4n+5)\mc{S}\Big)\e_{\a}h_{\a(n)}~.
\end{align}
\end{subequations}

Let us now consider the case when the on-shell superfield is type A or type B partially massless with super-depth $t$. Upon substituting $M=M^{(A)}_{(t,n)}$, \eqref{OS3b} reduces to
\begin{align}
0=\bigg(\mc{F}-\rho^{(-)}_{(t,n)}\bigg)h_{\a(n)}~,\qquad 0=\bigg(\mc{F}-\rho^{(-)}_{(t+1,n+1)}\bigg)\psi_{\a(n+1)}~.
\end{align}
On the other-hand, substituting $M=M^{(B)}_{(t,n)}$  we find 
\begin{align}
0=\bigg(\mc{F}-\rho^{(+)}_{(t,n)}\bigg)h_{\a(n)}~,\qquad 0=\bigg(\mc{F}-\rho^{(+)}_{(t,n+1)}\bigg)\psi_{\a(n+1)}~.~~~
\end{align}

We see that the type A supermultiplet $H_{\a(n)}^{(t,A)}$ consists of two negative helicity partially massless fields: one with spin $n/2$ and depth $t$, and the other with spin $(n+1)/2$ and depth $t+1$.  In contrast, the type B supermultiplet $H_{\a(n)}^{(t,B)}$ consists of two positive helicity depth-$t$ partially massless fields: one with spin $n/2$ and the other with spin $(n+1)/2$.

\subsection{Linearised higher-spin super-Cotton tensors}

In Minkowski superspace, the linearised higher-spin super-Cotton tensor \cite{K16,KT} is 
\bea
\mf{W}_{\a_1 \dots \a_n}(H) = 
\Big( -\frac{\ri}{2}\Big)^n
D^{\b_1} D_{(\a_1} \dots D^{\b_n} D_{\a_n)} H_{\b_1 \dots \b_n}~,
\label{5.17}
\eea
with $D_A = (\pa_a, D_\a)$ being the flat-superspace covariant derivatives.  
This tensor is invariant under the gauge transformations
\bea
\d_{\L} H_{\a(n) } &=& \ri^n D_{(\a_1 } \L_{\a_2 \dots \a_n)} ~, 
\label{5.18}
\eea
and obeys the conservation identity 
\bea
{D}^{\b} \mf{W}_{\b\a_1 \dots \a_{n-1}}(H)=0 ~.
\label{5.25}
\eea

In contrast to the non-supersymmetric case, explicit expressions for the higher-spin super-Cotton tensors are easily obtained in $\mc{N}=1$ AdS superspace. In \cite{KP18}, some of the lower rank super-Cotton tensors were derived using the operator
\begin{align}
\Delta^{\a}{}_{\b}:=-\frac{\text{i}}{2}\mc{D}^{\a}\mc{D}_{\b}-2\mc{S}\delta^{\a}{}_{\b} \label{Del1}
\end{align}
which satisfies the relations
\begin{align}
\mc{D}^{\b} \Delta^{\a}{}_{\b}=0~,\qquad \Delta^{\a}{}_{\b}\mc{D}_{\a}=0~. \label{ID82}
\end{align}
To derive the higher-spin super-Cotton tensors, we make use of the following extension of \eqref{Del1}
\begin{align}
\Delta_{[j]}^{\a}{}_{\b}:=-\frac{\text{i}}{2}\mc{D}^{\a}\mc{D}_{\b}-2j\mc{S}\delta^{\a}{}_{\b}~, \label{Del2}
\end{align} 
for which \eqref{Del1} corresponds to the $j=1$ instance, $\Delta^{\a}{}_{\b}\equiv \Delta_{[1]}^{\a}{}_{\b}$. Using the algebra \eqref{A8-mod}, it may be shown that they possess the following important properties
\begin{subequations}\label{DP}
\begin{align}
\big[\Delta_{[j]}^{\a_1}{}_{\b_1},\Delta_{[k]}^{\a_2}{}_{\b_2}\big]&=\ve_{\b_1\b_2}\mc{S}\big(\mc{D}^{\a(2)}-2\mc{S}M^{\a(2)}\big)-\ve^{\a_1\a_2}\mc{S}\big(\mc{D}_{\b(2)}-2\mc{S}M_{\b(2)}\big)~,\label{DP1}\\
\Delta_{[j]}^{\a}{}_{\g}\Delta_{[j+1]}^{\b}{}_{\d}\ve^{\d\g}&=j\mc{S}\ve^{\a\b}\big(\text{i}\mc{D}^2+4(j+1)\mc{S}\big)~,\label{DP2}\\
\Delta_{[j+1]}^{\a}{}_{\g}\Delta_{[j]}^{\b}{}_{\d}\ve_{\a\b}&=j\mc{S}\ve_{\g\d}\big(\text{i}\mc{D}^2+4(j+1)\mc{S}\big)~, \label{DP3}
\end{align}
\end{subequations}
for arbitrary integers $j$ and $k$.

In terms of \eqref{Del2}, the higher-spin super-Cotton tensor takes the remarkably simple form
\begin{align}
\mathfrak{W}_{\a(n)}(H)&=\Delta_{[1]}^{\b_1}{}_{(\a_1} \Delta_{[2]}^{\b_2}{}_{\a_2}\cdots\Delta_{[n]}^{\b_n}{}_{\a_n)}H_{\b(n)} \non\\
&=\Delta_{[n]}^{\b_1}{}_{(\a_1}\Delta_{[n-1]}^{\b_2}{}_{\a_2}\cdots\Delta_{[1]}^{\b_n}{}_{\a_n)}H_{\b(n)} \label{Scot}
\end{align}
The equivalence of the two above expressions for $\mathfrak{W}_{\a(n)}(H)$ follows from the identity \eqref{DP1}. The defining features of the super-Cotton tensors 
follow immediately from the properties \eqref{DP} of the operators \eqref{Del2}. In particular, transversality may be shown as follows
\begin{align}
\mc{D}^{\g}\mf{W}_{\g\a(n-1)}(H)&=\mc{D}^{\g}\Delta_{[1]}^{\b_1}{}_{(\g}\Delta_{[2]}^{\b_2}{}_{\a_1}\cdots\Delta_{[n]}^{\b_n}{}_{\a_{n-1})}H_{\b(n)} \non\\
&=\frac{1}{n!}\mc{D}^{\g}\big(\Delta_{[1]}^{\b_1}{}_{\g}\Delta_{[2]}^{\b_2}{}_{\a_1}\cdots\Delta_{[n]}^{\b_n}{}_{\a_{n-1}}+(n!-1) ~\text{permutations}~\big)H_{\b(n)}~. \label{scotT}
\end{align}
In the last line, all of the $(n!-1)$ permutations may be brought into the same form as the first term using \eqref{DP2}. From the first equation in \eqref{ID82}, it follows that the right hand side of  \eqref{scotT} vanishes, and hence
\begin{align}
0=\mc{D}^{\b}\mf{W}_{\b\a(n-1)}(H)~. \label{scotTT}
\end{align} 
In a similar vein, its variation under the gauge transformation \eqref{sgt}
may be computed as follows
\begin{align}
\mf{W}_{\a(n)}(\delta_{\L}H)&=\ri^n\Delta_{[n]}^{\b_1}{}_{(\a_1}\Delta_{[n-1]}^{\b_2}{}_{\a_2}\cdots\Delta_{[1]}^{\b_n}{}_{\a_n)}\mc{D}_{(\b_1}\L_{\b_2\dots\b_n)}\non\\
&=\frac{\ri^n}{n!}\Delta_{[n]}^{\b_1}{}_{(\a_1}\Delta_{[n-1]}^{\b_2}{}_{\a_2}\cdots\Delta_{[1]}^{\b_n}{}_{\a_n)}\big(\mc{D}_{\b_n}\L_{\b_1\dots\b_{n-1}}+(n!-1)~\text{permutations}~\big)~.
\end{align}
This time using \eqref{DP3}, all of the $(n!-1)$ permutations may be brought into the same form as the first term in the second line.  From the second identity in \eqref{ID82} it follows that the right hand side vanishes, 
\begin{align}
0=\mf{W}_{\a(n)}(\delta_{\L}H)~,
\end{align}
and hence $\mf{W}_{\a(n)}(H)$ is gauge invariant. 


\subsection{Factorisation of the higher-spin super-Cotton tensor}
  
The gauge invariance and transversality of the higher-spin super-Cotton tensor means that its associated higher-spin Chern-Simons-type functional \cite{KP18,KP19}
\begin{align}
 S^{(n)}_{\rm SCS}[H] = 
- \frac{\ri^n}{2^{\left \lfloor{n/2}\right \rfloor +1}} \int \rd^{3|2}z \, E \, H^{\a(n)} \mf{W}_{\a(n)}(H) ~, \qquad E^{-1}= {\rm Ber }(E_A{}^M)
\label{SCS}
\end{align}
is manifestly gauge invariant; here we have denoted $\rd^{3|2}z =\rd^3 x \rd^2\q$.
In the flat-superspace limit, the action \eqref{SCS} reduces to the one given in \cite{K16,KT}.
 On account of  the two equivalent forms of $\mf{W}_{\a(n)}(H)$ in \eqref{Scot}, it is also symmetric in the sense analogous to \eqref{sym}. 
 Since $ H^{\a(n)} $ and $\mf{W}_{\a(n)}(H) $ are primary superfields of dimensions
 \eqref{4.17} and \eqref{4.21} respectively, the action \eqref{SCS} is superconformal.

Similar to the non-supersymmetric case, we may impose the transverse gauge condition
\begin{align}
H_{\a(n)}\equiv H^{\text{T}}_{\a(n)}~,\qquad 0=\mc{D}^{\b}H^{\text{T}}_{\b\a(n-1)}~,
\end{align}
under which the super-Cotton tensor takes the form
\begin{align}
\mf{W}_{\a(n)}(H^{\text{T}})&=\frac{1}{(2n+1)^n}\prod_{t=0}^{ n-1 }\big(\mb{F}-M^{(A)}_{(t,n)}\big) H^{\text{T}}_{\a(n)} \non\\
&=\frac{1}{(2n+1)^n}\prod_{t=0}^{ \lceil n/2 \rceil -1 }\big(\mb{F}-M^{(A)}_{(t,n)}\big) \prod_{t=1}^{ \lfloor n/2 \rfloor }\big(\mb{F}-M^{(B)}_{(t,n)}\big) H^{\text{T}}_{\a(n)}~, \label{ScotFactor}
\end{align}
where we have used \eqref{ID9}. We see that on AdS${}^{3|2}$, the superconformal higher-spin action \eqref{SCS} factorises into first-order differential operators involving all of the type A and type B partial pseudo-mass values.  
  
Finally we note that, in any conformally flat superspace, it is also true that the vanishing of the higher-spin Super-Cotton tensor $\mf{W}_{\a(n)}(H)$ is a necessary and sufficient condition for 
$H_{\a(n)}$ to be pure gauge,
\begin{align}
\mathfrak{W}_{\a(n)}(H)=0 \quad \Longleftrightarrow \quad 
H_{\a(n)}=\ri^n\mc{D}_{\a}\L_{\a(n-1)}~,
\end{align}
 for some $\L_{\a(n-1)}$. This is discussed at the end of appendix \ref{AppendixC}, to where we refer the reader for more details. Therefore, from \eqref{ScotFactor}, it follows that both type A and B partially-massless superfields do not contain any local propagating degrees of freedom.

\subsection{Massive $\mc{N}=1$ gauge actions}

Off-shell $\mc{N}=1$ supersymmetric extensions of the new topologically massive higher-spin gauge models \eqref{HSNTMG} were also proposed in \cite{KP18}. For any integer $n\geq 1$, the gauge-invariant action corresponding to the unconstrained prepotential $H_{\a(n)}$ may be recast into the form 
\begin{align}
\mb{S}_{\text{NTM}}^{(n)}[H]=- \frac{\ri^n}{2^{\left \lfloor{n/2}\right \rfloor +1}}\frac{1}{M}\int \rd^{3|2}z \, E \,\mf{W}^{\a(n)}(H) \big(\mb{F}-M\big)H_{\a(n)}~. \label{SHSNTMG}
\end{align}
The equation of motion obtained by varying \eqref{SHSNTMG} with respect to $H_{\a(n)}$ is 
\begin{align}
0=\big(\mb{F}-M\big)\mf{W}_{\a(n)}(H)~. \label{SEOM1}
\end{align}
For generic $M$, this equation in conjunction with the off-shell conservation identity \eqref{scotTT}
means that the field-strength $\mf{W}_{\a(n)}(H)$ itself describes a propagating mode with pseudo-mass $M$, superspin $n/2$ and superhelicity $\frac{1}{2}\big(n+\frac{1}{2}\big)\s$, where $\s=M/|M|$. 
In the case when $M$ takes on one of the type A or type B partially massless values \eqref{SPM}, an analysis similar to that in the non-supersymmetric case can be conducted, with the conclusion that there are no local propagating degrees of freedom. 

Topologically massive supersymmetric higher-spin actions\footnote{These are higher-spin extensions of topologically massive $\mc{N}=1$ supergravity \cite{DK}.} involve different 
massless sectors depending on the value of superspin. For integer superspin, the relevant massless action\footnote{There is another massless action of this type, where the superfield $H_{\a(2s)}$ appears in the action with a first-order kinetic operator. However, the gauge transformations are of type B with $t=1$, $\delta_{\L}H_{\a(2s)}=\mc{D}_{\a(2)}\L_{\a(2s-2)}$. The corresponding action has been derived recently in \cite{HHK}.
 } is first-order (in vector derivatives) and takes the form \cite{KP18}
\bea
{\mathbb S}^{(2s)}_{\rm FO}[H,Y] &=& 
\bigg(-\frac{1}{2}\bigg)^s\frac{\ri}{2}
\int \rd^{3|2} z \, E\,\Big\{  
H^{\b  \a (2s-1)} \cD^\g \cD_\b H_{\g   \a (2s-1)}  \non \\
&&
 +2\ri (2s-1) Y^{\a (2s-2)} \cD^{\b(2)} H_{ \b(2) \a (2s-2) } \non \\
&&
+ (2s-1) \Big( Y^{\a (2s-2) }\cD^2 Y_{\a (2s-2)} 
+ 
(2s-2)\cD_\b Y^{\b \a (2s-3) }
\cD^\g Y_{\g \a (2s-3)} \Big) \non \\
&& -4\cS \ri \Big( H^{\a(2s) } H_{\a(2s) }
+2s(2s-1) Y^{\a(2s-2)} Y_{\a(2s-2)} \Big)
\Big\} ~.
\label{5.44}
\eea
It is invariant under the gauge transformations
\begin{subequations}
\bea
\d_{\L} {H}_{\a (2s)} &=& 
\cD_{\a } \L_{\a(2s-1)} ~, \\
\d_{\L} {Y}_{\a (2s-2) } &=& \frac{1}{2s} \cD^\b \L_{\b \a(2s-2)}~.
\eea
\end{subequations}
For half-integer superspin, the relevant massless action is second-order and takes the form  \cite{KP18}
\bea
{\mathbb S}_{\rm{SO}}^{(2s+1)}[H,Y]
&=& \bigg(- \hf \bigg)^s   \int \rd^{3|2} z \, E\,\bigg\{ 
-\frac{\ri}{2} H^{\a(2s+1)} {\mathbb Q} H_{\a(2s+1)}
-\frac{\ri}{8} \cD_\b H^{\b \a(2s)} \cD^2 \cD^\g  H_{\g \a(2s)} \non \\
&&+ \frac{\ri}{4} s \cD_{\b\g} H^{\b\g \a(2s-1)} 
\cD^{\r\l} H_{\r\l \a(2s-1)}
-\hf (2s-1) Y^{\a(2s-2)} \cD^{\b\g} \cD^\d H_{\b\g \d \a(2s-2)} \non \\
&&+\frac{\ri}{2}  (2s-1)\Big[ Y^{\a(2s-2)} \cD^2Y_{\a(2s-2)}
- \frac{s-1}{s} \cD_\b Y^{\b\a(2s-3)} \cD^\g Y_{\g \a(2s-3)}\Big]
\non \\
&& +\ri s \cS H^{\b \a(2s)}  \cD_\b{}^\g  H_{\g \a(2s)}
+\hf (s+1) {\cS} H^{\a(2s+1)} \cD^2 H_{\a(2s+1)}
 \label{5.45} \\
&&+\ri s (2s-3) {\cS}^2 H^{\a(2s+1)} H_{\a(2s+1)}
+ \frac{(2s-1)(s^2 -3s -2)}{s} {\cS} Y^{\a(2s-2)} Y_{\a(2s-2)}\bigg\}
~,~~~
\non
\eea
and it is invariant under the gauge transformations
\begin{subequations}
\bea
\d_{\L} H_{\a(2s+1)} &=& \ri \cD_{\a} \L_{\a(2s)} ~,\\
\d_{\L} Y_{\a(2s-2)} &=& \frac{s}{2s+1} \cD^{\b(2)} \L_{\b(2) \a(2s-2)}~.
\eea
\end{subequations} 
The corresponding massive gauge-invariant actions for integer and half-integer superspin are 
\begin{subequations} \label{5.39}
\bea
{\mathbb S}_{\rm TM}^{(2s)}[H,Y]
&=& {\mathbb S}_{\rm{SCS}}^{(2s)}[H]
+\boldsymbol{\mu}(M,s){\mathbb S}_{\rm{FO}}^{(2s)} [H,Y]~,
\label{5.39a}
\\
{\mathbb S}_{\rm TM}^{(2s+1)}[H,Y]
&=&{\mathbb S}_{\rm{SCS}}^{(2s+1)}[H]
+\boldsymbol{\nu}(M,s){\mathbb S}_{\rm{SO}}^{(2s+1)}[H,Y]~.
\label{5.39b}
\eea
\end{subequations}
The coupling constants $\boldsymbol{\mu}(M,s)$ and $\boldsymbol{\nu}(M,s)$ both have mass dimension $2s-1$, and are functions of $s$ and some parameter $M$ with mass dimension one. Their explicit form in terms of these quantities may be determined by requiring that any $H_{\a(n)}$ satisfying \eqref{OS2} is a particular solution to the resulting field equations, as in the non-supersymmetric case. We expect that both $\boldsymbol{\mu}$ and $\boldsymbol{\nu}$ will vanish at the type A and type B partially massless points \eqref{SPM}, where the models do not describe any local propagating degrees of freedom.  

The above massive gauge-invariant actions are manifestly supersymmetric, 
that is the $\cN=1$ AdS supersymmetry is realised off-shell. 
It is worth pointing out that there also exists an on-shell construction of gauge-invariant Lagrangian formulations for massive higher-spin ${\cN}=1$ supermultiplets in $\mathbb{R}^{2,1}$ and AdS$_{3}$ \cite{BSZ3, BSZ4, BSZ5}, extending previous works on the non-supersymmetric cases \cite{BSZ1, BSZ2}. These frame-like formulations are based on the gauge-invariant approach to the dynamics of massive higher-spin fields proposed by Zinoviev \cite{Zinoviev, Zinoviev2} and Metsaev \cite{Metsaev}. 


\section{Discussion} \label{section 5}

In this paper we have derived closed-form expressions for the Cotton tensors $\mf{C}_{\a(n)}(h)$ associated with the conformal higher-spin gauge fields $h_{\a(n)}$ 
  in AdS$_3$. In the fermionic (odd $n$) and bosonic (even $n$) cases, they are given by \eqref{cotF} and \eqref{cotB} respectively. Their properties proved to greatly facilitate the analysis of the on-shell dynamics of topologically massive higher-spin gauge models \eqref{TM}, and their new variant \eqref{HSNTMG}. In particular, it was shown that these models describe a propagating mode with pseudo mass $\rho$ and helicity sgn$(\rho)n/2$, except at the partially massless points \eqref{PM1}, where they describe only pure gauge degrees of freedom. 

In the transverse gauge, the Cotton tensors were demonstrated to factorise into products of second-order operators \eqref{CF} involving all partially massless values and, equivalently, into products of first-order operators \eqref{cotFO} involving all depth-$t$ pseudo-masses (with both signs of helicity). This in turn demonstrates that the action \eqref{CSA} for conformal higher-spin gauge fields factorises, on an AdS$_3$ background, into wave operators associated with partially massless fields. This is in line with earlier results obtained in even dimensions \cite{DeserN1,Tseytlin5,Tseytlin6,Tseytlin13,Karapet1,Karapet2,Metsaev:2014iwa,NTCHS,GH, KP20}. 
 
In Minkowski space, it is known \cite{BKLFP} that the higher-spin Cotton tensors are proportional to certain linear combinations of the positive and negative helicity spin-projection operators. The latter are the three-dimensional analogues of the Behrends-Fronsdal traceless and transverse (TT) projectors. Additionally, recent studies \cite{KP20} in AdS$_4$ have revealed that the poles of the corresponding TT projectors are intimately related to partially massless fields. This relation renders the factorisation properties of conformal higher-spin wave operators in AdS$_4$ transparent. It would be an interesting problem to construct the TT projectors in AdS$_3$ and investigate whether similar relations hold.  

In this paper we have also provided $\mc{N}=1$ AdS$_3$ superspace extensions to the non-supersymmetric results described above. Specifically, the remarkably simple expression \eqref{Scot} for the higher-spin super-Cotton tensor $\mf{W}_{\a(n)}(H)$ was derived. In the transverse gauge, it was shown that $\mf{W}_{\a(n)}(H)$ factorises into first-order wave operators \eqref{ScotFactor} associated with the so-called type A and type B partially massless superfields. Similar remarks  regarding the factorisation of the superconformal higher-spin gauge models \eqref{SCS}  in an AdS$_3$ superspace background follow. The fact that there are two types of partially massless supermultiplets, possessing gauge symmetries \eqref{SPMG1} and \eqref{SPMG2},  is a novel feature of three dimensions, and did not occur in four dimensions \cite{BHKP}. 

Finally, it would be interesting to obtain $\mc{N}$-extended analogues of the above results, as well as the corresponding AdS$_3$ superprojectors along the lines of \cite{BHHK, BHKP}.
\\

\noindent
{\bf Acknowledgements:}\\
SMK is grateful to Ulf Lindstr\"om and Gabriele Tartaglino-Mazzucchelli
for useful discussions. 
The authors are grateful to Daniel Hutchings for pointing out some notational 
inconsistencies and misprints in the manuscript. 
The work of SMK is supported in part by the Australian 
Research Council, project No. DP200101944.
The work of MP is supported by the Hackett Postgraduate Scholarship UWA,
under the Australian Government Research Training Program.

\appendix

\section{Two-component spinor toolkit} \label{appendix 0}

We follow the notation and conventions adopted in
\cite{KLT-M11}. In particular, the Minkowski metric is
$\eta_{ab}=\mbox{diag}(-1,1,1)$.
The spinor indices are  raised and lowered using
the $\rm SL(2,{\mathbb R})$ invariant tensors
\bea
\ve_{\a\b}=\left(\begin{array}{cc}0~&-1\\1~&0\end{array}\right)~,\qquad
\ve^{\a\b}=\left(\begin{array}{cc}0~&1\\-1~&0\end{array}\right)~,\qquad
\ve^{\a\g}\ve_{\g\b}=\d^\a_\b
\eea
by the standard rule:
\bea
\psi^{\a}=\ve^{\a\b}\psi_\b~, \qquad \psi_{\a}=\ve_{\a\b}\psi^\b~.
\label{A2}
\eea

We make use of real gamma-matrices,  $\g_a := \big( (\g_a)_\a{}^\b \big)$, 
which obey the algebra
\be
\gamma_a \gamma_b=\eta_{ab}{\mathbbm 1} + \varepsilon_{abc}
\gamma^c~,
\label{A3}
\ee
where the Levi-Civita tensor is normalised as
$\varepsilon^{012}=-\varepsilon_{012}=1$. 
Given a three-vector $V_a$,
it  can be equivalently described by a symmetric second-rank spinor $V_{\a\b}$
defined as
\bea
V_{\a\b}:=(\g^a)_{\a\b}V_a=V_{\b\a}~,\qquad
V_a=-\hf(\g_a)^{\a\b}V_{\a\b}~.
\eea
Any
antisymmetric tensor $F_{ab}=-F_{ba}$ is Hodge-dual to a three-vector $F_a$, 
specifically
\bea
F_a=\hf\ve_{abc}F^{bc}~,\qquad
F_{ab}=-\ve_{abc}F^c~.
\label{hodge-1}
\eea
Then, the symmetric spinor $F_{\a\b} =F_{\b\a}$, which is associated with $F_a$, can 
equivalently be defined in terms of  $F_{ab}$: 
\bea
F_{\a\b}:=(\g^a)_{\a\b}F_a=\hf(\g^a)_{\a\b}\ve_{abc}F^{bc}
~.
\label{hodge-2}
\eea
These three algebraic objects, $F_a$, $F_{ab}$ and $F_{\a \b}$, 
are in one-to-one correspondence to each other, 
$F_a \leftrightarrow F_{ab} \leftrightarrow F_{\a\b}$.
The corresponding inner products are related to each other as follows:
\bea
-F^aG_a=
\hf F^{ab}G_{ab}=\hf F^{\a\b}G_{\a\b}
~.
\label{A.7}
\eea

The Lorentz generators with two vector indices ($M_{ab} =-M_{ba}$),  one vector index ($M_a$)
and two spinor indices ($M_{\a\b} =M_{\b\a}$) are related to each other by the rules:
$M_a=\hf \ve_{abc}M^{bc}$ and $M_{\a\b}=(\g^a)_{\a\b}M_a$.
These generators 
act on a vector $V_c$ 
and a spinor $\J_\g$ 
as follows:
\bea
M_{ab}V_c=2\eta_{c[a}V_{b]}~, ~~~~~~
M_{\a\b}\J_{\g}
=\ve_{\g(\a}\J_{\b)}~.
\label{generators}
\eea
The following identities hold:
\begin{subequations}
\bea
M_{\a_1}{}^{\b}\F_{\b \a_2 ... \a_n} &=& - \hf (n+2)\F_{\a(n)}~,\\
M^{\b\g}M_{\b\g}  \F_{\a(n)} &=& -\hf n(n+2) \F_{\a(n)}
~.
\eea
\end{subequations}


\section{Generating function formalism} \label{appendix A}

In order to easily prove the defining properties of the higher-spin Cotton tensors, it will be useful to reformulate the problem into one in terms of homogeneous polynomials. This framework is also sometimes referred to as the generating function formalism. 

Associated with a
symmetric 
rank-$n$ spinor field $\phi_{\a(n)}:=\phi_{\a_1\dots\a_n}=\phi_{(\a_1\dots\a_n)}$ 
is 
 a homogeneous polynomial $\phi_{(n)}(\U)$ of degree $n$ defined by 
\begin{align}
\phi_{(n)}:=\U^{\a_1}\cdots\U^{\a_n}\phi_{\a_1 \dots \a_n}~,
\end{align}
where the  auxiliary variables $\U^{\a}$ are chosen to be commuting,
hence 
$\U^{\a}\U_{\a}:=\U^{\a}\ve_{\a\b}\U^{\b}=0$.
The correspondence $\phi_{\a(n)} \to \phi_{(n)}$ is one-to-one.
The linear space of such polynomials will be denoted $\mc{H}_{(n)}$.

Introducing the auxiliary derivative $\pa_{\a}:=\frac{\pa}{\pa\U^{\a}}$, whose index is raised according to the usual rule $\pa^{\a}:=\ve^{\a\b}\pa_{\b}$, we define the AdS differential operators
\begin{align}
\mc{D}_{(2)}:=\U^{\a}\U^{\b}\mc{D}_{\a\b}~,\qquad \mc{D}_{(0)}:=\U^{\a}\mc{D}_{\a}{}^{\b}\pa_{\b}~,\qquad \mc{D}_{(-2)}:=\mc{D}^{\a\b}\pa_{\a}\pa_{\b}~.
\end{align}
They raise the degree of homogeneity of any polynomial on which they act by 2, 0 and -2 respectively. Amongst themselves, they may be shown to satisfy the algebra\footnote{The auxiliary variables $\U^{\a}$ are defined to be inert with respect to the Lorentz generators. Alternatively, one could define their action on any $\phi_{(n)}\in \mc{H}_{(n)}$ to be $M_{\a\b}\phi_{(n)}:=-\U_{(\a}\pa_{\b)}\phi_{(n)}$.}
\begin{subequations}\label{ID2}
\begin{align}
\big[\mc{D}_{(2)}, \mc{D}_{(-2)}\big]&=4\big(\bm{\U}_{(0)} +1\big)\Box +8\mc{S}^2 \bm{M}_{(0)}\big(\bm{\U}_{(0)}+1\big)~,\label{ID2a}\\
\big[\mc{D}_{(2)}, \mc{D}_{(0)}\big]&=4\mc{S}^2\bm{M}_{(2)}\big(\bm{\U}_{(0)}+2\big)~,\label{ID2b}\\
\big[\mc{D}_{(-2)}, \mc{D}_{(0)}\big]&=-4\mc{S}^2\bm{M}_{(-2)}\big(\bm{\U}_{(0)}+2\big)~.\label{ID2c}
\end{align}
\end{subequations}
where we have made use of the definitions\footnote{The identities \eqref{ID3c} and \eqref{ID3d} imply that the commutators \eqref{ID2b} and \eqref{ID2c} vanish on $\mc{H}_{(n)}$. This is not surprising given that on $\mc{H}_{(n)}$, the operator $\mc{D}_{(0)}$ can be identified with the quadratic Casimir $\mc{F}$.}
\begin{subequations}\label{ID3}
\begin{align}
\bm{\U}_{(0)}&:=\U^{\a}\pa_{\a}~,\qquad \qquad ~~~~~~~~\bm{\U}_{(0)}\phi_{(n)}=n\phi_{(n)}~,\label{ID3a}\\
\bm{M}_{(0)}&:=\U^{\a}M_{\a}{}^{\b}\pa_{\b}~,\qquad \qquad~\bm{M}_{(0)}\phi_{(n)}=-\frac{1}{2}n(n+2)\phi_{(n)}~,\label{ID3b}\\
\bm{M}_{(2)}&:=\U^{\a}\U^{\b}M_{\a\b}~, \qquad \qquad \bm{M}_{(2)}\phi_{(n)}=0~,\label{ID3c}\\
\bm{M}_{(-2)}&:=M^{\a\b}\pa_{\a}\pa_{\b}~, \qquad \qquad \bm{M}_{(-2)}\phi_{(n)}=0~.\label{ID3d}
\end{align}
\end{subequations}
Using the above identities it is possible to show, via induction on $t$, that for any $\phi_{(n)}\in \mc{H}_{(n)}$ the following relations hold
\begin{subequations}
\begin{align}
\big[\mc{D}_{(2)}, \mc{D}^{\phantom{.}t}_{(-2)}\big]\phi_{(n)}&=4t(n-t+2)\big(\mc{Q}-\tau_{(t,n+2)}\mc{S}^2\big)\mc{D}_{(-2)}^{t-1}\phi_{(n)}~,\label{ID14}\\
\big[\mc{D}_{(-2)}, \mc{D}^{\phantom{.}t}_{(2)}\big]\phi_{(n)}&=-4t(n+t)\big(\mc{Q}-\tau_{(t,n+2t)}\mc{S}^2\big)\mc{D}_{(2)}^{t-1}\phi_{(n)}~.\label{ID15}
\end{align}
\end{subequations}
Here $\mc{Q}$ is the quadratic Casimir \eqref{Q} and $\tau_{(t,n)}$ are the partially massless values \eqref{PM2}.

\section{Gauge completeness} \label{AppendixC}

In this appendix we sketch a proof for the following result: In a conformally flat background $\cM^3$, 
the Cotton tensor $\mf{C}_{\a(n)}(h) $ vanishes if and only if the gauge field $h_{\a(n)}$ is pure gauge, 
\bea
\mf{C}_{\a(n)}(h) =0 \quad \Longleftrightarrow \quad 
h_{\a(n)}=\mc{D}_{\a(2)}\x_{\a(n-2)}~,
\label{C.1}
\eea
for some $\xi_{\a(n-2)}$. In the case of Minkowski space, this theorem was proved in \cite{HHL,HLLMP}.
Here we start by giving an alternative derivation in Minkowski space,
\bea
\mf{C}_{\a(n)}(h) =0 \quad \Longleftrightarrow \quad 
h_{\a(n)}=\pa_{\a(2)}\x_{\a(n-2)}~.
\label{C.1.5}
\eea

The Cotton tensor \eqref{Mcot} is invariant under the gauge transformations
\bea
\delta_{\x}h_{\a(n)}=\pa_{\a(2)}\x_{\a(n-2)}~.
\label{C.2}
\eea
This gauge freedom allows us to choose the gauge 
\bea
\pa^{\b(2) } h_{\b(2) \a(n-2)} = 0~,
\label{C.3}
\eea
in which the Cotton tensor takes the form \cite{KP18}
\begin{subequations}
\bea
\mf{C}_{\alpha(2s)}(h)&=&\Box^{s-1}\partial^{\beta}{}_{(\a_1}h_{\a_2\dots\a_{2s})\beta}
=\Box^{s-1}\partial^{\beta}{}_{\a_1}h_{\a_2\dots\a_{2s}\beta}
~, 
\label{C.4a}
\\
\mf{C}_{\alpha(2s+1)}(h)&=&\Box^s h_{\alpha(2s+1)}~.
\label{C.4b}
\eea
\end{subequations}
Now suppose $\mf{C}_{\a(n)}(h) =0 $. We assume $h_{\a(n)}$ to decrease at infinity, 
such that its Fourier transform is well defined. Then it follows that for $n>2$ the prepotential obeys the wave equation 
\bea
\Box h_{\a(n) } =0~,
\label{C.5}
\eea
in addition to the transverse condition \eqref{C.3}. The equations \eqref{C.3} and \eqref{C.5} are invariant under a restricted class of gauge transformations \eqref{C.2}
such that the gauge parameter $\x_{\a(n-2)} $ is constrained by 
\bea
\pa^{\b(2) } \x_{\b(2) \a(n-4)} = 0~, \qquad \Box \x_{\a(n-2)} = 0~.
\eea
This residual gauge freedom suffices to gauge $h_{\a(n)}$ away, which implies 
\eqref{C.1.5}. The case $n=2$ is special since the condition $\mf{C}_{\alpha(2)}(h)=0$
gives the first-order equation
\bea
\partial^{\beta}{}_{(\a_1} h_{\a_2) \beta}
=\partial^{\beta}{}_{\a_1}h_{\a_2\beta}=0 \quad \implies \quad \Box h_{\a(2) } =0~,
\eea
and thus the vector field $h_{\a(2)}$ has a single independent massless component. 
The residual gauge freedom is described by a parameter $\x$ constrained by 
$\Box \x =0$. This gauge freedom allows us to gauge away $h_{\a(2)}$. 

To complete the proof of \eqref{C.1}, it remains to point out that both equations in \eqref{C.1} are invariant under Weyl transformations in curved space 
\bea
\cD_a \to \cD'_a =  \re^{\s} \big( \cD_a +\cD^b\s M_{ba}\big)~,
\eea
provided $h_{\a(n)} $ and $\x_{\a(n-2)} $ are primary fields of dimension
$(2-n/2 )$ and $(1-n/2)$, respectively. As discussed in \cite{KP18}, the Weyl transformation law of $h_{ \a(n)}$ implies that $\mf{C}_{\a(n)}(h) $ is a primary 
field of dimension \eqref{1.9}.

Yet another way to arrive at the result \eqref{C.1.5} in Minkowski space is by using the 3D spin-projection operators of \cite{BKLFP}. The latter are defined according to
\begin{align}
\Pi^{(\pm)}_{~\a}{}^{\b}=\frac{1}{2}\bigg(\delta_{\a}{}^{\b}\pm\frac{\pa_{\a}{}^{\b}}{\sqrt{\Box}}\bigg)~,\qquad \Pi^{(\pm n)}_{~\a(n)}{}^{\b(n)}=\Pi^{(\pm )}_{~(\a_1}{}^{\b_1}\dots \Pi^{(\pm)}_{~\a_n)}{}^{\b_n}~,
\end{align}
and satisfy the relations
\begin{align}
 \Pi^{(+n)}  \Pi^{(+n)} = \Pi^{(+n)} ~, \qquad 
 \Pi^{(-n)}  \Pi^{(-n)} = \Pi^{(-n)} ~, \qquad 
 \Pi^{(+n)}  \Pi^{(-n)} = 0~.
\end{align} 
Let us define the following combinations
\begin{align}
\Pi_{\perp}^{[n]}:=\Pi^{(n)}+\Pi^{(-n)}~,\qquad \Pi_{\parallel}^{[n]}:=\mathds{1}-\Pi_{\perp}^{(n)}~.\label{C.11}
\end{align} 
It is clear that they resolve the identity, $\mathds{1}=\Pi_{\perp}^{[n]}+ \Pi_{\parallel}^{[n]}$, and are orthogonal projectors
\begin{align}
\Pi_{\perp}^{[n]}  \Pi_{\perp}^{[n]}=\Pi_{\perp}^{[n]} ~, \qquad 
 \Pi_{\parallel}^{[n]}  \Pi_{\parallel}^{[n]}= \Pi_{\parallel}^{[n]} ~, \qquad 
\Pi_{\perp}^{[n]}  \Pi_{\parallel}^{[n]} = 0~.
\end{align}
Furthermore, $\Pi_{\perp}^{[n]}$ projects a field $h_{\a(n)}$ onto its transverse component,
\begin{align}
h^{\perp}_{\a(n)}&\equiv \Pi_{\perp}^{[n]}h_{\a(n)}~,\qquad  ~~~~~0=\pa^{\b(2)}h^{\perp}_{\a(n-2)\b(2)}~,
\end{align}
 whilst $\Pi_{\parallel}^{[n]}$ projects onto its longitudinal (pure gauge) component,
\begin{align}
h^{\parallel}_{\a(n)}&\equiv \Pi_{\parallel}^{[n]}h_{\a(n)}~,\qquad  h^{\parallel}_{\a(n)}=\pa_{\a(2)}\xi_{\a(n-2)}~,
\end{align}
for some $\xi_{\a(n-2)}$. It follows that any unconstrained field $h_{\a(n)}$ may be decomposed according to
\begin{align}
h_{\a(n)}=h^{\perp}_{\a(n)}+\pa_{\a(2)}\xi_{\a(n-2)}~.
\end{align}
To arrive at the conclusion \eqref{C.1.5}, it remains to recall that the higher-spin Cotton tensors can be expressed as 
\begin{align} 
\mathfrak{C}_{\a(2s)}(h)=\Box^{s-1} \Pi_{\perp}^{[2s]}\pa_{\a}{}^{\b}h_{\a(2s-1)\b}~,\qquad \mathfrak{C}_{\a(2s+1)}(h)&=\Box^{s}  \Pi_{\perp}^{[2s+1]} h_{\a(2s+1)}~.
\end{align}
These relations imply that 
\bea
h^{\perp}_{\a(2s)}=\Box^{-s} \pa_{\a}{}^{\b}{\mathfrak C}_{\a(2s-1)\b}(h)~,\qquad 
h^{\perp}_{\a (2s+1)}=\Box^{-s} \mathfrak{C}_{\a(2s+1)}(h)~,
\eea
which completes the proof of  \eqref{C.1.5}.

In \cite{BHHK}, the superprojectors in $\mc{N}$-extended Minkowski superspace were derived, and expressions for the corresponding higher-spin linearised super-Cotton tensors were given. The above argument may be replicated in a straightforward manner using the results of \cite{BHHK}. The conclusion is that, in any conformally-flat superspace, a necessary and sufficient condition for a prepotential to be pure gauge is if its super-Cotton tensor vanishes.


\begin{footnotesize}

\end{footnotesize}

\end{document}